\documentclass[%
prd,
 superscriptaddress,
preprintnumbers,
nofootinbib,
 amsmath,amssymb,
]{revtex4-1}

\usepackage{amsmath}
\usepackage{amssymb}
\usepackage{amsthm}
\usepackage{amsfonts}
\usepackage{natbib}

\usepackage{graphicx}

\usepackage[normalem]{ulem}

\usepackage{dcolumn}
\usepackage{bm}

\usepackage{color}
\usepackage[usenames,dvipsnames,svgnames,table]{xcolor}
\usepackage{slashed,euscript}
\newcommand{\bea}{\begin{eqnarray}}
\newcommand{\eea}{\end{eqnarray}}
\newcommand{\be}{\begin{equation}}
\newcommand{\ee}{\end{equation}}

\begin{document}

\title{Status of background-independent coarse-graining in tensor models for quantum gravity}
 
 \author{Astrid Eichhorn}
   \email{a.eichhorn@thphys.uni-heidelberg.de}
\affiliation{Institut f\"ur Theoretische
  Physik, Universit\"at Heidelberg, Philosophenweg 16, 69120
  Heidelberg, Germany}
  
  \author{Tim Koslowski}
  \email{koslowski@nucleares.unam.mx}
  \affiliation{Instituto de Ciencias Nucleares, UNAM,
Apartado Postal 70-543, Coyoac\'an 04510, Ciudad de M\'exico, Mexico}

 \author{Antonio D.~Pereira}
   \email{a.pereira@thphys.uni-heidelberg.de}
\affiliation{Institut f\"ur Theoretische
  Physik, Universit\"at Heidelberg, Philosophenweg 16, 69120
  Heidelberg, Germany}

\begin{abstract}
A background-independent route towards a universal continuum limit in discrete models of quantum gravity proceeds through a background-independent form of coarse graining. This review provides a pedagogical introduction to the conceptual ideas underlying the use of the number of degrees of freedom as a scale for a Renormalization Group flow. We focus on tensor models, for which we explain how the tensor size serves as the scale for a background-independent coarse-graining flow.
This flow provides a new probe of a universal continuum limit in tensor models. We review the development and setup of this tool and summarize results in the 2- and 3-dimensional case. Moreover, we provide a step-by-step guide to the practical implementation of these ideas and tools by deriving the flow of couplings in a rank-4-tensor model. We discuss the phenomenon of dimensional reduction in these models and find tentative first hints
for an interacting fixed point with potential relevance for the continuum limit in four-dimensional quantum gravity.
\end{abstract}

\pacs{Valid PACS appear here}

\maketitle

\section{Invitation to  background-independent coarse-graining in tensor models  for quantum gravity}
 The path-integral for quantum gravity takes center stage in a diverse range of approaches to quantum spacetime. It is tackled either as a quantum field theory for the metric \cite{Hawking:1979ig,Donoghue:1994dn,Reuter:1996cp,Reuter:2012id,Eichhorn:2018yfc,Feldbrugge:2017kzv},  or in a discretized fashion with a built-in regularization \cite{Hamber:2009zz,Perez:2012wv,Ambjorn:1998xu,Ambjorn:2001cv,
 Ambjorn:2012jv,Laiho:2016nlp,Rivasseau:2011hm,Rivasseau:2012yp,Rivasseau:2013uca,Rivasseau:2016zco,Hamber:2009mt,Freidel:2005qe,Baratin:2011hp}. The latter approach, relying on unphysical building blocks of space(time), 
 provides access to a physical space(time) only when a universal continuum limit can be taken. Universality \cite{Cardy:1996xt,ZinnJustin:2002ru,Rosten:2010vm} is key in this setting, as it guarantees independence of the physics from unphysical choices, e.g., in the discretization procedure, i.e., the shape of the building blocks. To discover universality, background-independent coarse-graining techniques are a well-suited tool  as universality arises at fixed points of the coarse-graining procedure.
 
The notion of ``background independent coarse-graining" at a first glance appears to be an oxymoron and suggests this review should be extremely short. After all, in order to coarse grain, one first needs to define what one means by ``coarse" and by ``fine". 
Intuitively one would expect these notions to rely on a background. In particular, a definition of ultraviolet and infrared, key to the setup of Renormalization Group (RG) techniques, seems to require a metric, i.e., a geometric background.
Yet, with RG techniques now playing an important role in different quantum-gravity approaches, 
coarse-graining techniques suitable for a setting without distinguished background have successfully been developed
\cite{Bahr:2010cq,Dittrich:2011zh,Dittrich:2012jq,Dittrich:2013bza,Dittrich:2013xwa,Dittrich:2014ala,Bahr:2014qza,Becker:2014qya,Morris:2016spn,Dittrich:2016tys,Delcamp:2016dqo,Bahr:2016hwc,Bahr:2017klw,Bahr:2018gwf,Lang:2017beo,Eichhorn:2018akn} and applied to various quantum-gravity models. 
In this review, we will 
focus on
the developments kicked off in 
\cite{Brezin:1992yc,Eichhorn:2013isa,Eichhorn:2014xaa,Gurau:2010ba,Rivasseau:2011hm,Eichhorn:2013isa}, and
introduce the key concepts behind a 
 background-independent RG flow and the associated notion of coarse graining.  In particular, we will focus on the development and application of these tools to tensor models. 
 
\begin{figure}[!b]
\includegraphics[scale=.7]{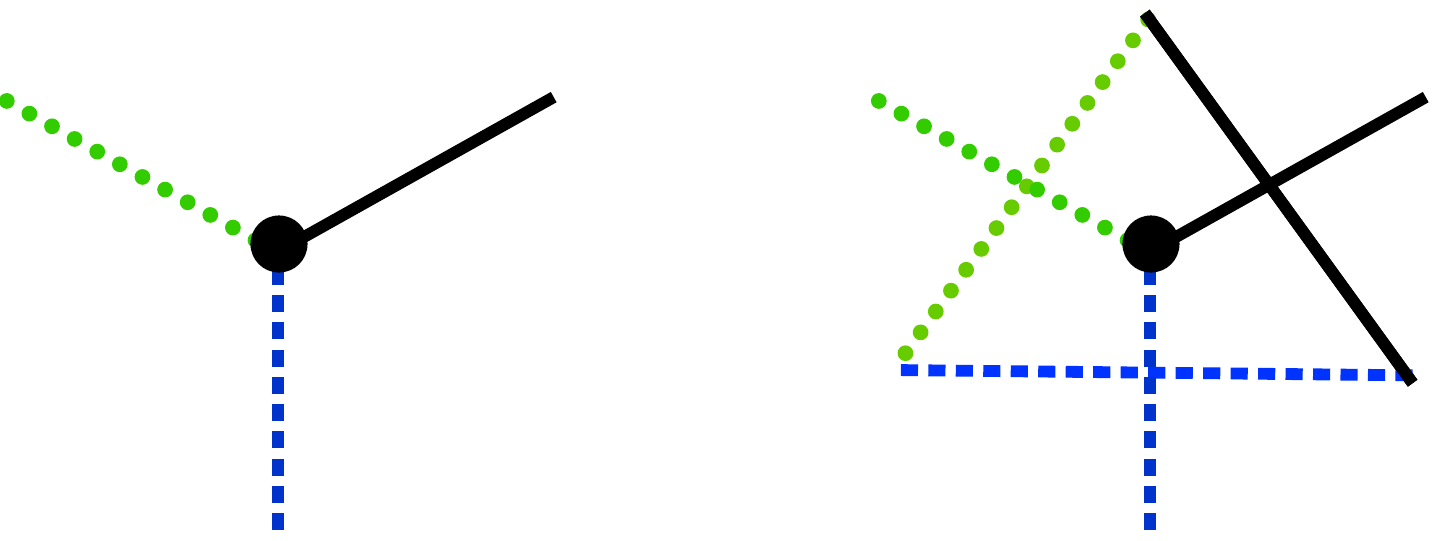}
\caption{\label{fig:tensors3}  The three indices of a rank-3-tensor are associated to the three lines of a triangle.}
\end{figure}

Tensor models are of interest for quantum gravity both as a way of exploring the partition function\cite{Rivasseau:2011hm,Gurau:2011xp,Gurau:2016cjo,Bonzom:2016dwy} directly as well as through a conjectured correspondence of specific tensor models to aspects of a geometric description 
in the context of the SYK-model \cite{Witten:2016iux,Klebanov:2016xxf,Gurau:2016lzk}. In both settings, the large $N'$ limit, where $N'$ is the tensor size, is of key interest, and physical results are extracted in the limit $N' \rightarrow \infty$.
In their simplest version that is of particular interest to quantum gravity, tensor models are 0-dimensional theories, i.e., there is no notion of spacetime in the definition of the models. Instead, the \emph{dual} interpretation of the interactions in tensor models is that of discrete building blocks of space(time), cf.~Fig.~\ref{fig:tensors3}.  In this interpretation of tensor models through the graphs dual to the Feynman diagrams, the building blocks are interpreted as pieces of flat space(time). Curvature is accordingly localized at the hinges
\footnote{Closely related to tensor models are tensorial (group) field theories which  
are characterized by
the same non-trivial combinatorial structure of tensor models interactions. In addition, a non-trivial kinetic term is present, and group field theories are defined on a
group manifold, see, e.g., \cite{BenGeloun:2012yk,Geloun:2013saa,Geloun:2012bz,BenGeloun:2011rc,BenGeloun:2012pu,Samary:2014tja,Carrozza:2012uv,Carrozza:2013wda}. This extra group data may be associated with intrinsic geometric data of the associated simplices.}. The dual representation of tensors is in terms of building blocks of geometry.

\begin{figure}[!t]
\includegraphics[scale=.7]{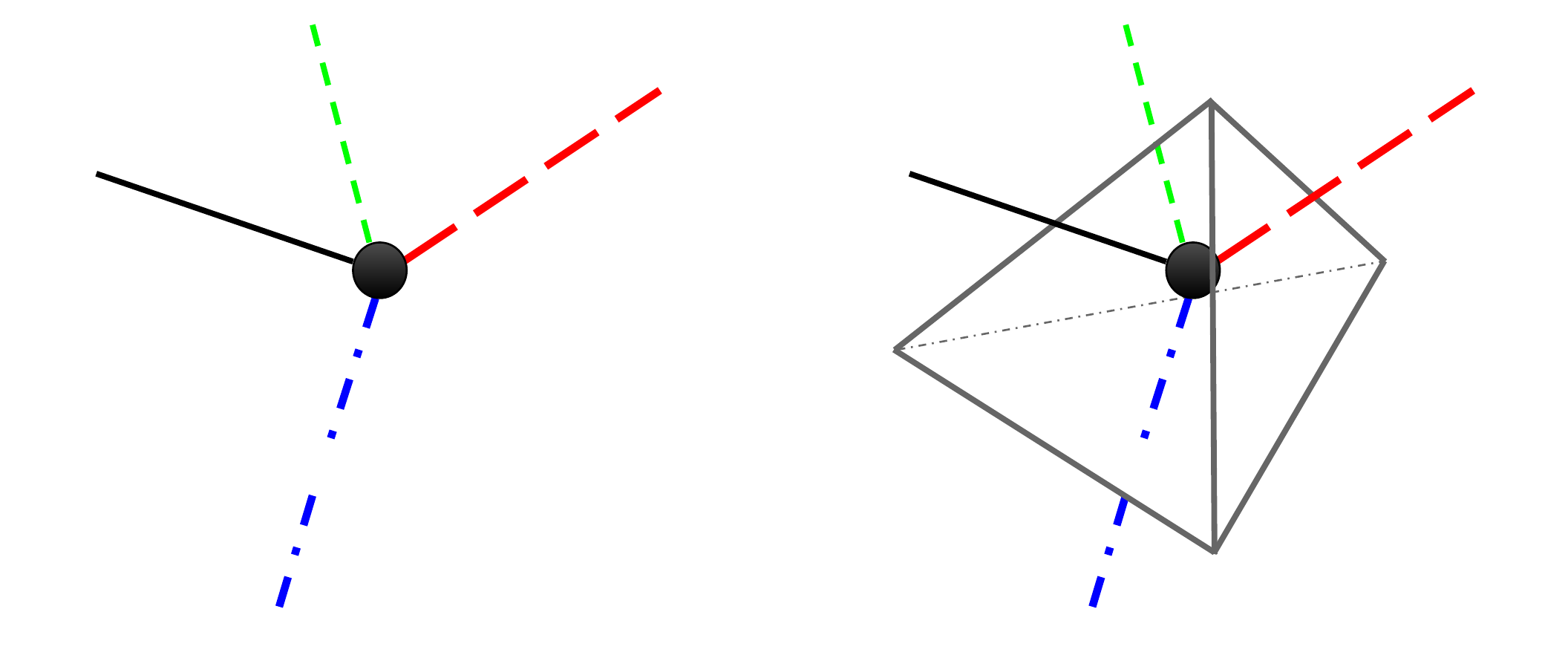}
\caption{\label{fig:tensors2}  The four indices of a rank-4-tensor are associated to the three triangles of a tetrahedron.}
\end{figure}

\begin{figure}[!t]
\includegraphics[scale=.5]{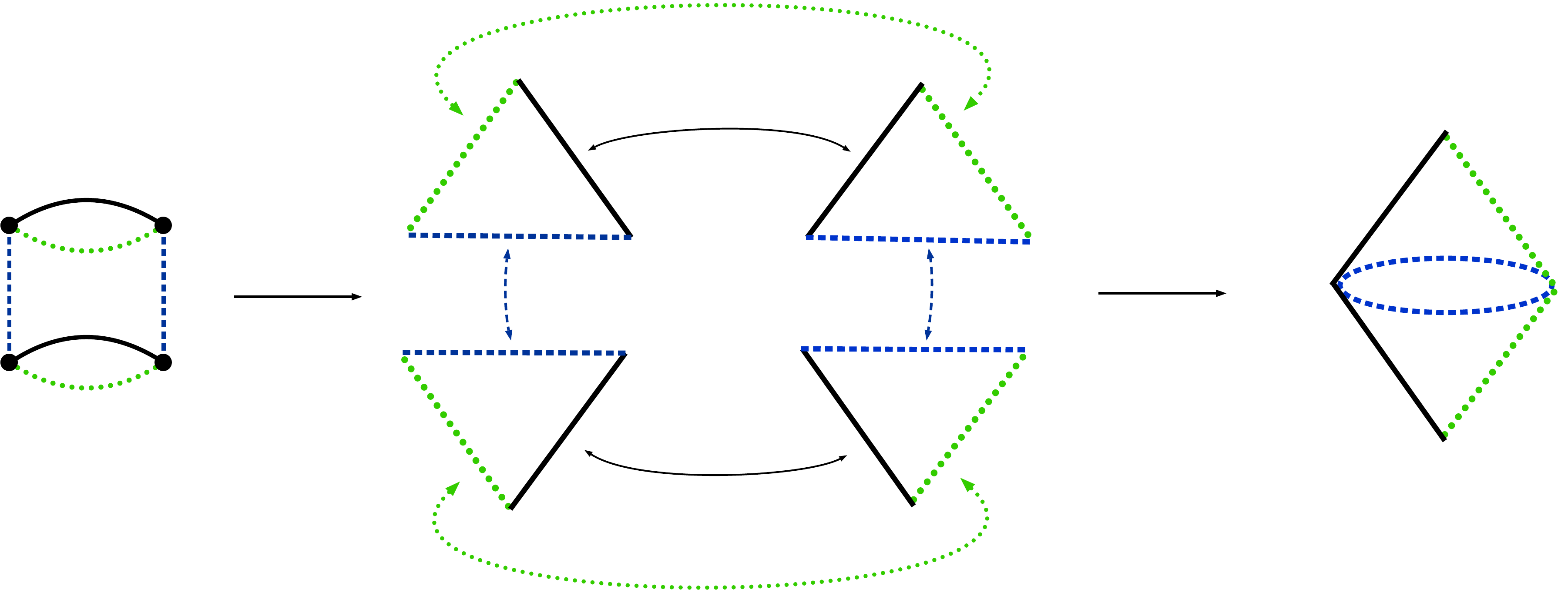}
\caption{\label{fig:tensors4} The invariant $T_{ijk}T_{ijl}T_{mnl}T_{mnk}$, depicted on the left, is associated to the gluing of four triangles (center) into a building block of 3-space (to the right). The contraction of common indices is associated to the gluing of triangles along common edges.}
\end{figure}

The $d$ indices of a rank-$d$ tensor are associated to the $(d-2)$- subsimplices of a $(d-1)$-simplex. For instance, for rank 3, the indices are associated to the edges ($(d-2)$ subsimplices) of a triangle ($(d-1)$ simplex), cf.~Fig.~\ref{fig:tensors3}. Correspondingly, in the rank-4-case, each index is associated to one of the four faces of a tetrahedron, cf.~Fig.~\ref{fig:tensors2}.
When two tensors are contracted along one index, the corresponding $(d-1)$-simplices share a $(d-2)$ simplex, e.g., for the rank-3 case, two triangles are glued along an edge, cf.~Fig.~\ref{fig:tensors4}. In the rank-4-case, two tetrahedra are glued along a face.
Allowed interaction terms are positive powers of the tensors that contain no free indices. This means that each $(d-2)$- subsimplex is glued to another $(d-2)$- subsimplex.
Therefore they correspond to $d$-dimensional building blocks of space(time), e.g., tetrahedra for a fourth-order interaction in the rank-3 case, cf.~Fig.~\ref{fig:tensors3}. The propagator of the theory identifies all $d$ indices of two tensors, corresponding to a glueing of one $d-1$ simplex to another, e.g., glueing of two triangles along their faces. Accordingly, the terms in the Feynman diagram expansion of tensor models have a dual interpretation as simplicial pseudomanifolds. In other words, the combinatorics of tensor models encode dynamical triangulations. In the simplest case, when no additional rules are imposed on the gluing, Riemannian pseudomanifolds are generated. The inscription of local lightcones inside the building blocks, such that a consistent notion of causality can emerge and the pseudomanifold is Lorentzian, requires additional rules for the gluing and more than one type of building block \cite{Ambjorn:1998xu,Ambjorn:2001cv,Ambjorn:2012jv,Jordan:2013awa}.

There are no experimental hints that indicate that spacetime is a simplicial pseudo-manifold,  accordingly it is assumed to be a continuum manifold.  In particular, while the presence of physical discreteness close to the Planck scale could be compatible with all observations to date, one would not expect a naive discretization as it arises from tensor models, to actually be physical. Instead, this form of discreteness should be regarded merely as a regularization of the path integral.
In order to 
take the continuum limit in tensor models, the number of degrees of freedom, encoded in the tensor size $N'$, must be taken to infinity. In \cite{Gurau:2009tw,Gurau:2010ba,Gurau:2011xp,Gurau:2011aq,Gurau:2011xq,Bonzom:2012hw,Carrozza:2015adg} it was shown that models of real (complex) tensors with a $O(N')\otimes O(N')\otimes...\otimes O(N')$ ($U(N')\otimes U(N')\otimes...\otimes U(N')$) symmetry\footnote{Here we use the notation $G \otimes G$ to denote the direct product of group actions as linear transformations of different indices of tensors.} admit a $1/N'$ expansion, where $N'$ is the size of the tensors. Here, each symmetry group in the above product acts on exactly one of the indices of the tensor. 
Due to the existence of a $1/N'$ expansion,
these are viable candidates to search for a physical continuum limit by taking $N'\rightarrow \infty$. Yet, simply taking $N'\rightarrow \infty$ is not sufficient in order to obtain a physical continuum limit: The microscopic properties and structure of the building blocks in the model is not taken to be physical, but only a discretization/regularization. Different microscopic choices can be made that should not leave an imprint on the continuum physics, such as, e.g., the shape of the building blocks. Accordingly, the continuum limit should be \emph{universal}. Universality is achieved at fixed points of the Renormalization Group flow. Therefore, an RG flow must be set up for these models. Unlike in quantum field theories defined on a background, no local, i.e., geometric
notion of scale is available. In fact, the only notion of scale is the size of the tensors, $N'$. 
In fact, using the tensor size $N'$ as a scale agrees with the intuitive notion of coarse graining, also underlying formal developments such as the a-theorem \cite{Cardy:1988cwa}: Coarse-graining leads from many degrees of freedom (large $N'$), to fewer, effective degrees of freedom (small $N'$).
Therefore, a pregeometric RG flow is set up in the tensor size $N'$, where a universal continuum limit can then be discovered as an RG fixed point. This point of view was advocated in \cite{Brezin:1992yc,Rivasseau:2011hm} 
and formally developed and benchmarked in \cite{Eichhorn:2013isa,Eichhorn:2014xaa,Eichhorn:2017xhy,Eichhorn:2018ylk}. 

In the dual picture, the lattice spacing needs to be taken to zero in such a way that the correlation length on the lattice diverges. Then, microscopic details of the setup become irrelevant. This is possible at a higher-order phase transition, linked to a fixed point in the space of couplings. The intuition behind these ideas can be tested in the two-dimensional case, where the double-scaling limit of matrix models \cite{Douglas:1989ve,Brezin:1990rb,Gross:1989vs,Gross:1989aw}, which is a universal continuum limit, is obtained by taking $N'\rightarrow \infty$ while tuning the coupling to a critical value as a power of $N'$. This is completely analogous to the case of continuum RG flows, where universal critical behavior with diverging correlation length is tied to RG fixed points, near which couplings scale with particular powers of the scale. Specifically, the double-scaling limit in matrix models with coupling $g$ is achieved by taking $N'\rightarrow \infty$ and $g\rightarrow g_{\rm crit}$, while holding 
\be
(g-g_{\rm crit})^{\frac{5}{4}}N' =\rm const,
\ee
which can be rewritten in the form
\be
g(N')= g_{\rm crit} + {\textrm{const.}^{4/5}\,}N^{-4/5}.
\ee
This immediately brings to mind the linearized scaling of couplings close to RG fixed points, which is given by the scale raised to the power $-\theta$, with the critical exponent $\theta$.

Note that there are arguments suggesting that quantum gravity should be discrete. One might interpret this as implying that there is no need to take the continuum limit in tensor models, and one can instead even work at finite $N'$. Yet, discreteness is actually a subtle issue in quantum gravity.
As discussed in more detail, e.g., in \cite{Eichhorn:DICE},
kinematical and dynamical discreteness are not the same thing in quantum gravity, and discreteness can be an \emph{emergent} property of the physical continuum limit. On the other hand, a simple implementation of discreteness in the sense of a cutoff potentially features the same breakdown of predictivity at scales near the cutoff that effective field theories do. Specifically, the interaction terms compatible with the symmetries of a model are  infinitely many for tensor models. The continuum limit is a way of imposing predictivity in a model by reducing the number of free parameters characterizing its dynamics to finitely many.  In the  RG language, this is linked to the fact that fixed points feature only finitely many relevant directions. 
In the language of critical phenomena, one has to tune only finitely many parameters to approach criticality in the sense of a higher-order phase transition. In this spirit, we aim at discovering a universal continuum limit in tensor models for quantum gravity such that both independence of unphysical microscopic details as well as predictivity is guaranteed. We leave open the question whether these models feature \emph{emergent} discreteness once the continuum limit is taken, but merely point out that taking the continuum limit does in fact not preclude the possibility of emergent, physical discreteness.

In summary, to discover a universal continuum limit, at which a physical spacetime could emerge from discrete building blocks of spacetime, we must discover universal critical points. These are linked to  RG fixed points. In the absence of a background, the only scale available for coarse-graining is the tensor size $N'$. As we will explain in the next sections, setting up an RG flow in $N'$ is both conceptually meaningful as well as feasible in practice.

This review is structured as follows. In Sec.~\ref{sec:basics} we introduce the conceptual basics of background-independent coarse graining. We provide an overview of how to implement these ideas in practice and how to 
set up
a flow equation in Sec.~\ref{sec:FRG}. In Sec.~\ref{sec:Scalingdim} we discuss in detail how scaling dimensions can be derived in a setting without a background, translating to the absence of physical length scales and corresponding units that would define canonical dimensions. We provide an overview of the benchmark case of two dimensions in Sec.~\ref{sec:matrixmodel}, where quantitatively robust results on the well-known continuum limit can be achieved using our flow equation. In Sec.~\ref{sec:rank3} we summarize results in the rank-3-case, where several RG fixed points give access to a dimensionally reduced continuum limit. We also highlight a recently discovered candidate for a fixed point
which might potentially turn out to be relevant
for three-dimensional quantum gravity. To provide a step-by-step instruction in how to set up and evaluate RG flows in tensor models, we present the first study of a rank-4-model with these tools in Sec.~\ref{sec:rank4}. We discover several universality classes featuring dimensional reduction. As a hint of the promise our method could have, we unveil tentative indications for a universality class that 
might potentially
be linked to four-dimensional quantum gravity.
In the outlook and conclusions \ref{sec:conclusions} we advocate that progress towards a comprehensive understanding of quantum gravity could be accelerated by strengthening the effort to bridge the gap between different approaches to quantum gravity. We discuss in particular how continuum studies of asymptotic safety, Monte Carlo simulations of (causal) dynamical triangulations and FRG studies of tensor models could provide a link to phenomenology and particle physics, while allowing to probe features of emergent geometries and enabling us to link the discrete and continuum side via a universal transition.

\section{Conceptual basics: Background independent Renormalization Group flow in gravity}\label{sec:basics}
Renormalization Group techniques are playing a role in several different approaches to quantum gravity. This includes
the asymptotic-safety program \cite{Reuter:2012id,Eichhorn:2018yfc}, 
the continuum limit in spin foams \cite{Bahr:2010cq,Dittrich:2011zh,Dittrich:2012jq,Dittrich:2013bza,Dittrich:2013xwa,Dittrich:2014ala,Bahr:2014qza,Dittrich:2016tys,Delcamp:2016dqo,Bahr:2016hwc,Bahr:2017klw,Bahr:2018gwf} and Hamiltonian RG flows in canonical loop quantum gravity \cite{Lang:2017beo},  tensorial (group) field theories \cite{BenGeloun:2011rc,Carrozza:2016vsq} as well as
holographic RG flows in the context of the AdS/CFT conjecture \cite{Skenderis:2002wp}. Yet, at a first glance, quantum gravity would appear to be the one of the fundamental interactions to which RG techniques are not easily applicable. The reason lies in the dichotomy of background independence and local coarse graining. While the results obtained with a local coarse-graining formulation can be made background independent, see, e.g., \cite{Becker:2014qya}, the RG flow itself necessarily relies on an (auxiliary) background, if the flow has the interpretation of a local coarse graining. A more direct reconciliation of RG techniques with background independence is provided by a nonlocal form of coarse-graining: RG flows, in agreement with the a-theorem \cite{Cardy:1988cwa}, connect descriptions with many degrees of freedom with effective descriptions of the same system based on fewer degrees of freedom. This idea can be realized both in a local as well as a nonlocal form. The latter is directly applicable to tensor models for quantum gravity. These are defined without any notion of spacetime, metric or locality. Yet they come with a measure of the number of degrees of freedom, namely the tensor size $N'$. Coarse-graining therefore corresponds to integrating out subsequent ``layers" of the tensors (rows and columns in the matrix-model case), thereby connecting a description at large $N'$ with an effective description at small $N'$. In particular, such coarse-graining techniques  
allow us to search for a well-defined large $N'$-limit, where the dynamics stays invariant under the step from $N'$ to $N'+1$, such that the limit $N' \rightarrow \infty$ can be taken. In this limit, one can hope for quantum space(time) to emerge from tensor models.\\
Note also that while local coarse-graining techniques typically rely on Riemannian signature, raising the difficulty of connecting back to the Lorentzian case of interest for physics, a nonlocal coarse graining does not rely on a momentum cutoff. Accordingly, a more direct search for a universal continuum limit for Lorentzian models could become possible in this setup. This includes applications of the FRG to tensor models dual to causal dynamical triangulations \cite{Benedetti:2008hc} as in \cite{Castro:FRGCDT}, as well as the application of coarse-graining techniques to the link matrix in causal sets \cite{Eichhorn:2017bwe}.

We will now explain how to implement these ideas in practice in the form of a flow equation.
One can view the flow equation as a reformulation of the path integral in terms of a functional differential equation. The search for a continuum limit in the path integral then becomes the search for a well-defined ultraviolet (in an appropriate sense) solution of the flow equation. At a completely general and formal level, the derivation of the flow equation from the path integral works as follows: One introduces a new term into the exponential in the generating functional, that is quadratic in the field and depends on some external parameter which we will call $\mathcal{K}$ here. For now, we leave this parameter completely general, and do not provide any physical interpretation associated with it. It is simply to be thought of as a ``sieve" on the space of field configurations, letting through only a subset of configurations.
The generating functional depends on $\mathcal{K}$ and is denoted by $Z_{\mathcal{K}}$, schematically
\be
Z_{\mathcal{K}} = \int \mathcal{D}\varphi\, e^{-S[\varphi]+ {\rm Tr}J\cdot \varphi - \frac{1}{2}{\rm Tr}\varphi\cdot R_{\mathcal{K}}\cdot \varphi},\label{eq:defZK}
\ee
where $S[\varphi]$ is a given microscopic action, $J$ is an external source and $\varphi$ denotes the random fields. The trace is to be interpreted in a suitable way for the model at hand, i.e., it signifies a momentum integral and trace over internal indices in standard QFTs on a background, and an appropriate summation over indices in the discrete case, e.g., for tensor models. 
We do not write indices for simplicity, but the fields are not necessarily scalars. As a function of 
the
parameter $\mathcal{K}$, a subset of configurations in the generating functional are suppressed, such that in the limit $\mathcal{K} \rightarrow \infty$, all configurations are suppressed. Conversely,
 in the limit $\mathcal{K}\rightarrow 0$ the unmodified generating functional is recovered. Since the suppression term is quadratic in the field, $\partial_{\mathcal{K}}Z_{\mathcal{K}}$ 
 can be expressed in terms of
 the two-point function,
\be
\partial_{\mathcal{K}}Z_{\mathcal{K}} =-\frac{1}{2}  \int \mathcal{D}\varphi\,{\rm Tr}\,\varphi\cdot (\partial_{\mathcal{K}}R_{\mathcal{K}})\cdot \varphi\, e^{-S[\varphi]+ {\rm Tr}J\cdot \varphi- \frac{1}{2}{\rm Tr}\varphi\cdot R_{\mathcal{K}}\cdot \varphi}.
\ee
For the modified Legendre transform
\be
\Gamma_{\mathcal{K}}[\phi] = \underset{J}{\rm sup} \left({\rm Tr}J\cdot \phi - \ln Z_{\mathcal{K}} \right)-  \frac{1}{2}{\rm Tr}\phi\cdot R_{\mathcal{K}}\cdot \phi,
\ee
with $\phi = \langle \varphi \rangle$, this implies
\be
\partial_{\mathcal{K}} \Gamma_{\mathcal{K}}[\phi]  = \frac{1}{2}{\rm Tr} \left[\left(\frac{\delta^2 \Gamma_{\mathcal{K}}[\phi] }{\delta \phi^2}+ R_{\mathcal{K}} \right)^{-1}\partial_{\mathcal{K}}R_{\mathcal{K}}\right],\label{eq:floweqgeneral}
\ee
which is known as the functional renormalization group (FRG) equation. For the case of a continuum QFT on an (auxiliary) background it was derived in  \cite{Wetterich:1992yh}, see also
  \cite{Ellwanger:1993mw,Morris:1993qb}, pioneered for gauge theories in \cite{Reuter:1993kw} and gravity in \cite{Reuter:1996cp}.
Up to here, the derivation of the flow equation from the path integral is just a formal ``trick" that can be performed with any (functional) integral: Instead of performing the integral ``all at once", one introduces the exponential of a quadratic term that depends on an external parameter. 
This allows to
 derive a differential equation that encodes how the result of the integral reacts to changes in the parameter.  As long as the suppression term 
 is quadratic in the field, an equation which is structurally of the form Eq.~\eqref{eq:floweqgeneral} follows directly from the definition Eq.~\eqref{eq:defZK}. 
 The question to address in a physics setting is whether any physical meaning can be given  
to the external parameter and consequently to the ensuing differential equation. \\
For instance,
in  
local field theories  
introducing an external parameter that does not lead to a notion of local coarse graining is not expected to be fruitful. In such cases, the modes that remain after integrating out some ``shells" of modes do not contain physically relevant degrees of freedom.  Thus deriving effective field theories for those degrees of freedom might be an interesting computational exercise, but is presumably not useful for answering physical questions. The notion of UV/IR is therefore key to make the effective field theories obtained by renormalization useful for practical computations. 
Thus, although
both in QFTs with and without a background, different choices for $\mathcal{K}$ are possible, ``non-local" choices have not yet been tested for their usefulness in the setting with a background.
Accordingly, in the case with a background it turns out to be the most powerful tool to  
 relate $\mathcal{K}$ to a momentum scale. This choice allows to implement a notion of local coarse graining: Decomposing configurations into eigenfunctions of an appropriate Laplacian,  $R_{\mathcal{K}}$ suppresses configurations with eigenvalues of the Laplacian smaller than $k^2$. In this case, the flow equation has the interpretation of 
providing
 the response of the effective dynamics to a local coarse-graining step. 
 \\
 The quest for a well-defined path integral, which exists as all configurations are taken into account  becomes the question for a well-defined solution of Eq.~\eqref{eq:floweqgeneral} for $\mathcal{K} \rightarrow \infty$. Specifically, in tensor models, it is useful to choose the suppression term as a function of the number of components of the tensor, $N$, e.g., in the form
\be
\Delta S_N = \frac{1}{2}{\rm Tr}\,{T_{a_1...a_d}R_N(a_1,...,a_d)T_{a_1...a_d}},
\ee 
such that
\be
\partial_t \Gamma_N[T]=N \partial_N \Gamma_N[T] = \frac{1}{2}{\rm Tr}\left[\left(\frac{\delta^2\Gamma_N}{\delta T_{a_1...a_d}\delta T_{b_1...b_d}}+ R_N(a_1...a_d)\delta_{a_1b_1}...\delta_{a_db_d} \right)^{-1} \partial_{t}R_N(a_1...a_d)\right].\label{eq:floweqtensors}
\ee
In a slight abuse of notation, we use $T$ both for the tensors that are integrated over in the generating functional, as well as for their expectation value on which the effective average action $\Gamma_N$ depends.

As we search for a phase transition in these models, we will employ the FRG to search for infrared (IR) fixed points.  
The relevant directions correspond to the number of parameters that require tuning to reach criticality. As we aim at approaching such IR fixed points in the limit of large tensors, we will set up beta functions in the large $N$ limit.

The general structure of the flow equation was first derived and benchmarked
in the case of matrix models in \cite{Eichhorn:2013isa,Eichhorn:2014xaa} and  
applied and further developed for
rank-3 tensor models in \cite{Eichhorn:2017xhy,Eichhorn:2018ylk}. Similarly, the FRG has been employed in the context of tensorial (group) field theories in \cite{Benedetti:2014qsa,Benedetti:2015yaa,Geloun:2015qfa,Geloun:2016qyb,Lahoche:2016xiq,Carrozza:2016tih,Geloun:2016xep,Carrozza:2016vsq,Carrozza:2017vkz,BenGeloun:2018ekd,Lahoche:2018oeo,Lahoche:2018ggd}. See also \cite{Krajewski:2015clk,Krajewski:2016svb} for related studies using the Polchinski equation. 

In the following, we will review how to use Eq.~\eqref{eq:floweqtensors} for practical calculations and to search for candidates for a universal continuum limit in quantum gravity.

\section{Lightning review of the setup: Theory space, regulator and how to calculate in practice}\label{sec:FRG}
\subsection{Theory space}
The flow equation provides the scale-dependent change of coefficients of the dynamics, spanned by the infinitely many terms compatible with the symmetries.
For instance starting with a quartic interaction term in $\Gamma_N [T]$, Eq.~\eqref{eq:floweqtensors} takes the schematic form
\be
\partial_t \Gamma_N[T] \sim \frac{\#}{\#+T^2},
\ee
which admits a Taylor expansion with nonvanishing coefficients not only of the quartic but generically also of all $T^{2n}$, $n \in \mathbb{N}$. The is the analogue of the well-known observation that Wilsonian coarse-graining flow generates all quasi-local interactions that are compatible with the symmetries \footnote{In the local case, the well-known non-locality of the full effective action $\Gamma_{k\rightarrow 0}$ is expected to arise through resummation of quasi-local terms, i.e., terms with arbitrary high but positive powers of derivatives, see, e.g., \cite{Codello:2010mj}. If nonlocal terms, i.e., terms with negative powers of derivatives and/or fields are included in theory space as independent basis elements, predictivity is expected to break down, even at interacting fixed points, as these interaction terms have increasingly positive canonical dimension. Although there is no notion of quasilocality in spacetime in the tensor model interactions, terms with negative powers of tensors are expected to suffer from the same problem. Moreover, terms with inverse powers of tensors do not directly provide an interpretation in terms of building blocks of geometry in a dual picture.
In fact, if a ``quasilocal" truncation of theory space is chosen, no interactions which cannot be written in a quasilocal form are generated by the flow at finite scales. Accordingly, this restriction of the theory space is both well-motivated as well as self-consistent.}. 
Even though in the case at hand we are not dealing with a local coarse-graining flow, the analogous observation holds and all interactions with positive powers of tensors that obey the  symmetries, are generated. Accordingly, 
to implement the flow equation in practice requires
 the following steps
\begin{itemize}
\item[a)] understanding which interactions are part of the (infinite-dimensional) theory space,
\item[b)] selecting a criterion according to which truncations of the theory space to a (finite-dimensional) subspace can be 
 chosen,
\item[c)] truncating theory space to a subspace in which Eq.~\eqref{eq:floweqtensors} can be evaluated in practice,
\item[d)] finding solutions of Eq.~\eqref{eq:floweqtensors} and checking whether they satisfy the criterion in b).
\end{itemize}
Steps c) and d) are then iterated and only fixed-point solutions which reach stability under the steps in the iteration procedure are kept.

The tensor models\footnote{In this review, we call tensor models 0-dimensional theories of random tensors. The kinetic term is a product of Kronecker deltas, i.e, there is no non-trivial kinetic operator which breaks the $O(N^\prime)^{\otimes d}$ (or $U(N^\prime)^{\otimes d}$) symmetry.} typically of interest for quantum gravity feature an independent symmetry group for each index. position, e.g., a product of $d$ copies of an $O(N')$ symmetry for the real rank-$d$ model. Accordingly, interactions cannot have an explicit index dependence, and no tensors with open indices can occur. All
allowed interactions $\mathcal{O}^{(n)}$ of $n$ tensors can therefore be 
cast
in the form
\be
\mathcal{O}^{(n)} = T_{a_1 b_1...d_1}...T_{a_n b_n...d_n} \mathcal{C}_{a_1...a_n, b_1...b_n,...,d_1...d_n},\label{eq:invariants}
\ee
where $d$ is the rank. The contraction pattern $ \mathcal{C}_{a_1...a_n, b_1...b_n,...,d_1...d_n}$ is a product of Kronecker deltas, in which $a$'s can only be contracted with $a$'s, $b$'s with $b$'s and so forth. All possible permutations of the labels 1 to 
 $n$ have to be taken into account independently for each index set $a$, $b$, etc. Some of the resulting $\mathcal{O}^{(n)}$ will be combinatorially equivalent, in which case only one representative is taken into account. In Table~\ref{tablecombstruct} we list combinatorially distinct structures up to sixth order in the tensors for the real and complex rank-3 models.
 
\begin{table}[!ht]
\centering
\begin{tabular}{|c||c||c||c|} 
\hline
\begin{tabular}[c]{@{}c@{}}number \\ of \\tensors \end{tabular} &  invariant & \begin{tabular}[c]{@{}c@{}}graphical \\representation  \\ (real model)\end{tabular} & \begin{tabular}[c]{@{}c@{}}graphical \\representation  \\ (complex model)\end{tabular}  \\ \hline \hline
2  & $T_{abc}T_{abc}$ & \begin{minipage}{3cm}
		\begin{center}
			\includegraphics[scale=.3]{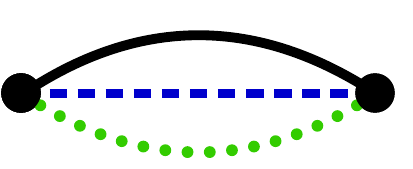}
		\end{center}
	\end{minipage}& \begin{minipage}{3cm}
		\begin{center}
			\includegraphics[scale=.3]{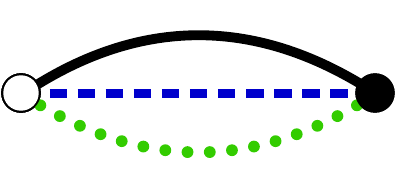}
		\end{center}
	\end{minipage}
\\ \hline
4 &  $T_{a_1 a_2 a_3}T_{b_1 a_2 a_3} T_{b_1 b_2 b_3} T_{a_1 b_2 b_3}$ & \begin{minipage}{3cm}
		\begin{center}
			\includegraphics[scale=.3]{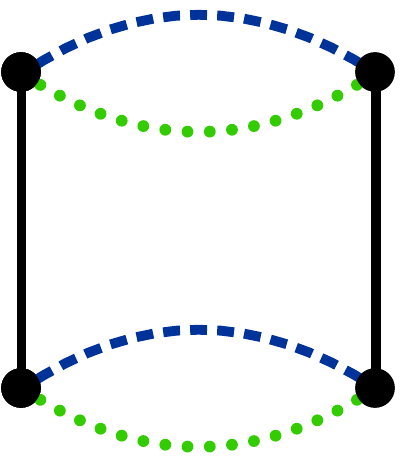}\\
		\end{center}
	\end{minipage}& \begin{minipage}{3cm}
		\begin{center}
			\includegraphics[scale=.3]{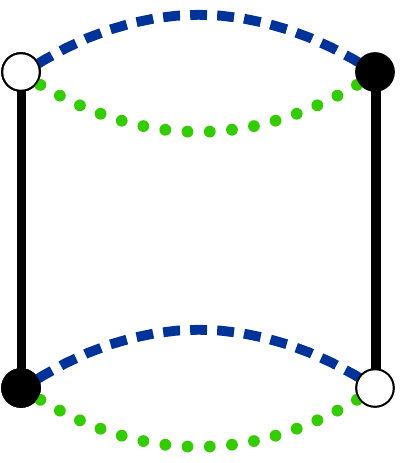}\\
		\end{center}
	\end{minipage}
\\ \hline
4 &  $T_{a_1 a_2 a_3}T_{a_1 b_2 b_3} T_{b_1 a_2 b_3} T_{b_1 b_2 a_3}$ & \begin{minipage}{3cm}
		\begin{center}
			\includegraphics[scale=.3]{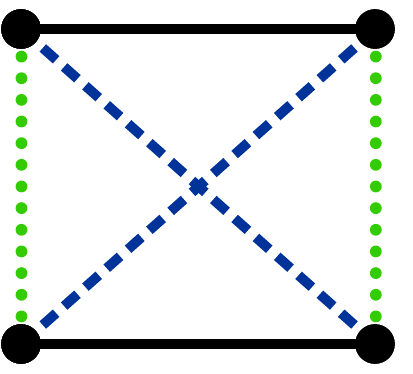}\\
		\end{center}
	\end{minipage} & $-$
\\ \hline
4 &  $T_{a_1 a_2 a_3}T_{a_1 a_2 a_3} T_{b_1 b_2 b_3} T_{b_1 b_2 b_3}$ & \begin{minipage}{3cm}
		\begin{center}
			\includegraphics[scale=.3]{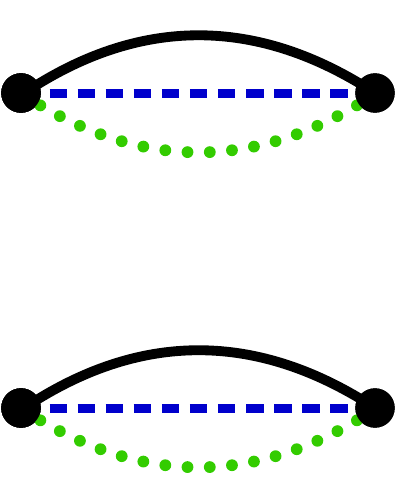}\\
		\end{center}
	\end{minipage} & \begin{minipage}{3cm}
		\begin{center}
			\includegraphics[scale=.3]{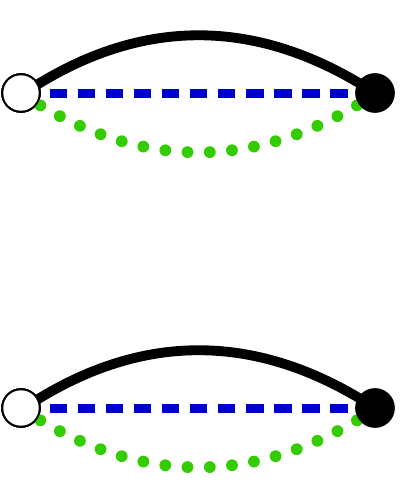}\\
		\end{center}
	\end{minipage}
\\ \hline
6 &  $T_{a_1 a_2 a_3}T_{b_1 a_2 a_3} T_{b_1 b_2 b_3} T_{c_1 b_2 b_3}T_{c_1 c_2 c_3} T_{a_1 c_2 c_3}$ & \begin{minipage}{3cm}
		\begin{center}
			\includegraphics[scale=.3,angle =90]{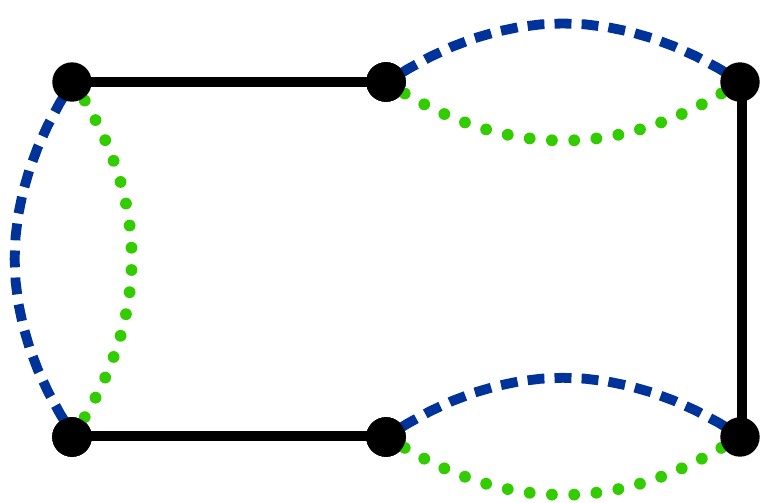}\\
		\end{center}
	\end{minipage} & \begin{minipage}{3cm}
		\begin{center}
			\includegraphics[scale=.3,angle =90]{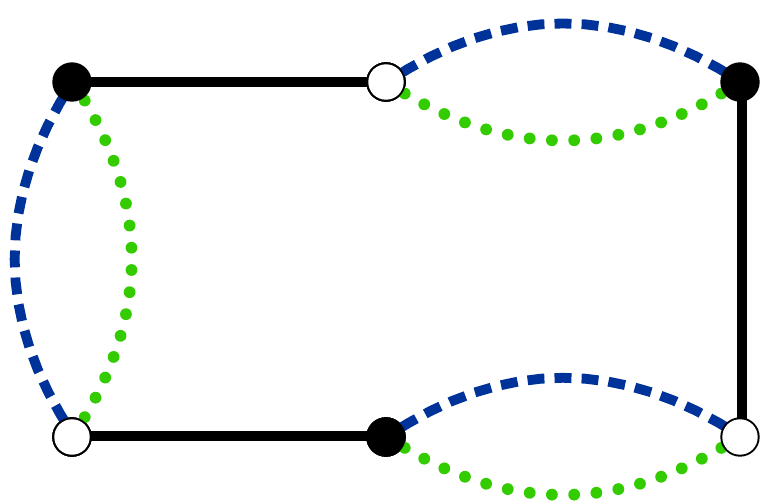}\\
		\end{center}
	\end{minipage}
\\ \hline
6 &  $T_{a_1 a_2 a_3} T_{a_1 b_2 a_3} T_{b_1 b_2 b_3} T_{c_1 c_2 b_3} T_{c_1 c_2 c_3} T_{b_1 a_2 c_3}$ & \begin{minipage}{3cm}
		\begin{center}
			\includegraphics[scale=.3,angle =90]{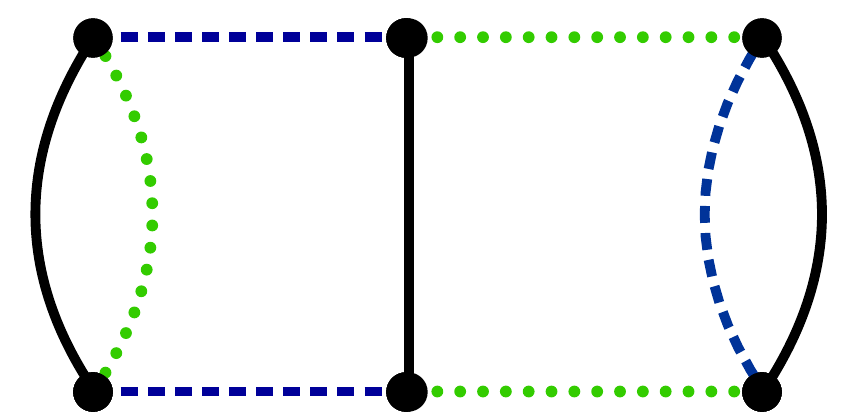}\\
		\end{center}
	\end{minipage} & \begin{minipage}{3cm}
		\begin{center}
			\includegraphics[scale=.3,angle =90]{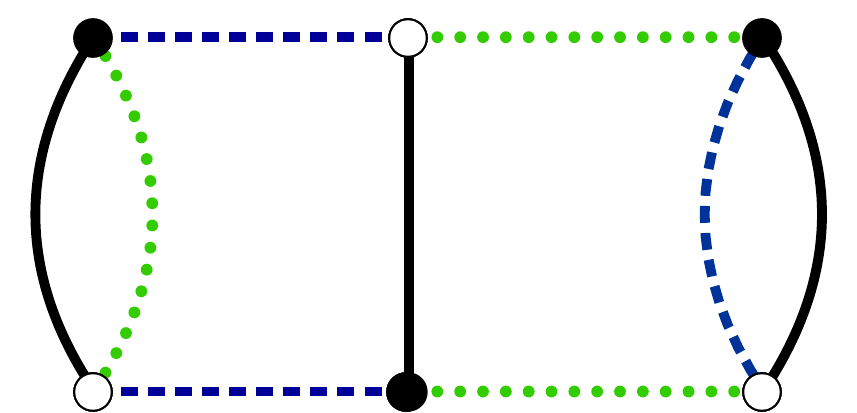}\\
		\end{center}
	\end{minipage}
\\ \hline
6 &  $T_{a_1 a_2 a_3} T_{a_1 b_2 b_3} T_{b_1 b_2 a_3} T_{b_1 c_2 c_3} T_{c_1 c_2 c_3} T_{c_1 a_2 b_3}$ & \begin{minipage}{3cm}
		\begin{center}
			\includegraphics[scale=.3,angle =90]{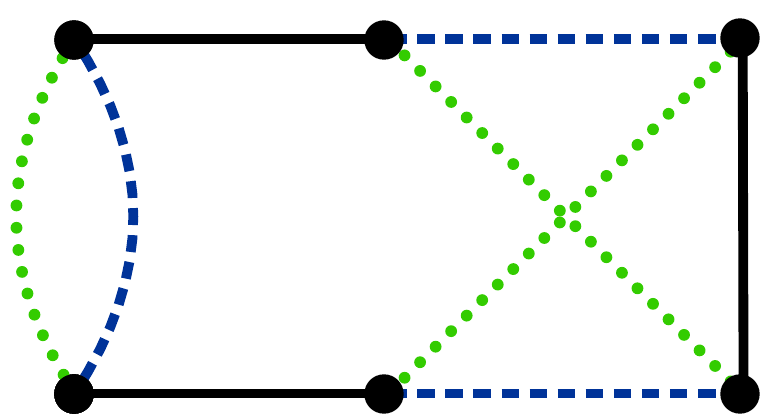}\\
		\end{center}
	\end{minipage} & $-$
\\ \hline
6 &  $T_{a_1 a_2 a_3} T_{b_1 a_2 b_3} T_{b_1 c_2 c_3} T_{c_1 c_2 a_3} T_{c_1 b_2 b_3} T_{a_1 b_2 c_3}$ & \begin{minipage}{3cm}
		\begin{center}
			\includegraphics[scale=.3,angle =90]{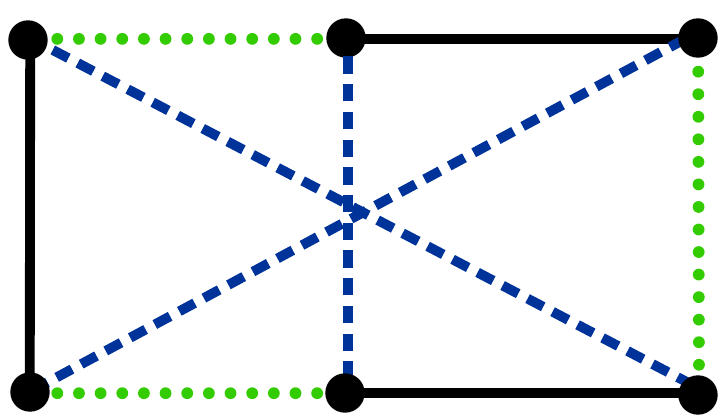}\\
		\end{center}
	\end{minipage}  & \begin{minipage}{3cm}
		\begin{center}
			\includegraphics[scale=.3,angle =90]{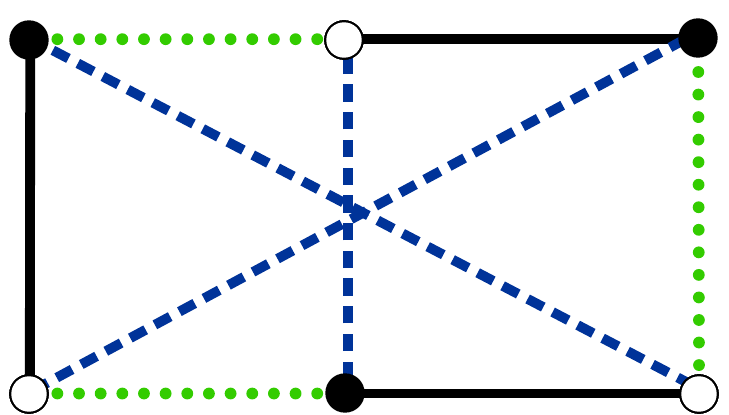}\\
		\end{center}
	\end{minipage}
\\ \hline
6 &  $T_{a_1 a_2 a_3}T_{b_1 a_2 b_3} T_{b_1 b_2 c_3}T_{c_1 b_2 b_3}T_{c_1 c_2 a_3}T_{a_1 c_2 c_3}$ & \begin{minipage}{3cm}
		\begin{center}
			\includegraphics[scale=.3,angle =90]{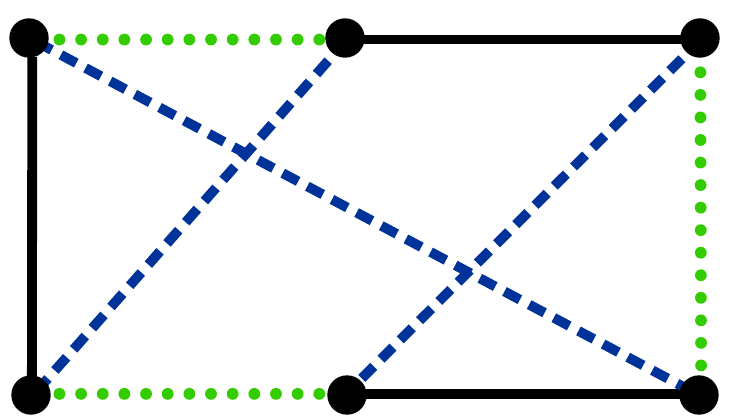}\\
		\end{center}
	\end{minipage} & $-$
\\ \hline
6 &  $T_{a_1 a_2 a_3}T_{a_1 a_2 a_3} T_{b_1 b_2 b_3}T_{b_1 b_2 b_3}T_{c_1 c_2 c_3}T_{c_1 c_2 c_3}$ & \begin{minipage}{3cm}
		\begin{center}
			\includegraphics[scale=.3,angle =90]{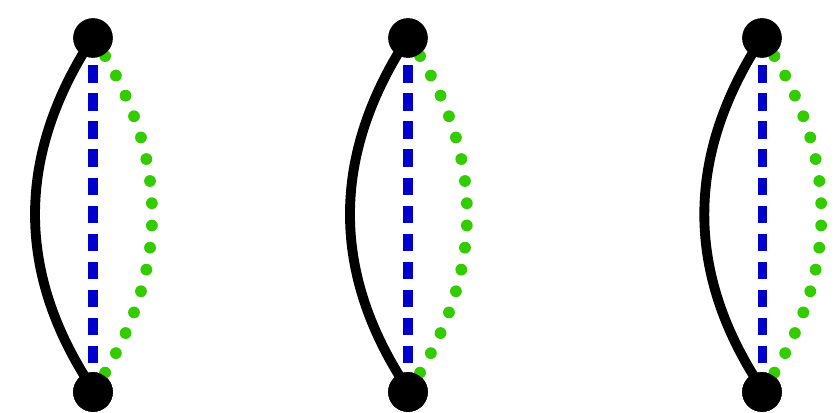}\\
		\end{center}
	\end{minipage} & \begin{minipage}{3cm}
		\begin{center}
			\includegraphics[scale=.3,angle =90]{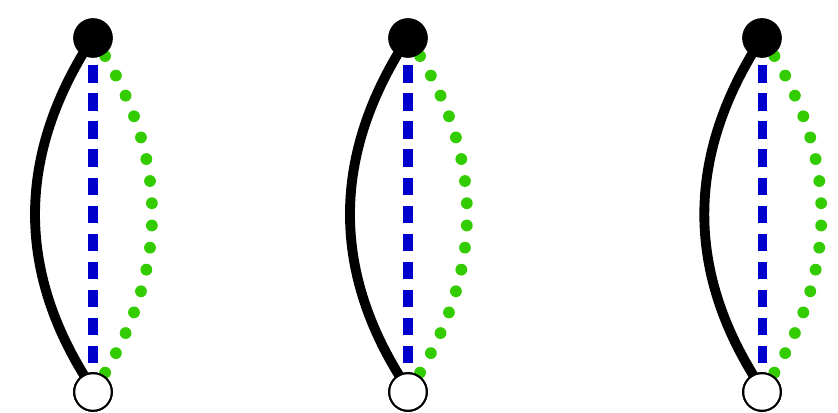}\\
		\end{center}
	\end{minipage}
\\ \hline
6 &  $T_{a_1 a_2 a_3} T_{b_1 a_2 a_3} T_{b_1 b_2 b_3} T_{a_1 b_2 b_3} T_{c_1 c_2 c_3} T_{c_1 c_2 c_3}$ & \begin{minipage}{3cm}
		\begin{center}
			\includegraphics[scale=.3,angle =90]{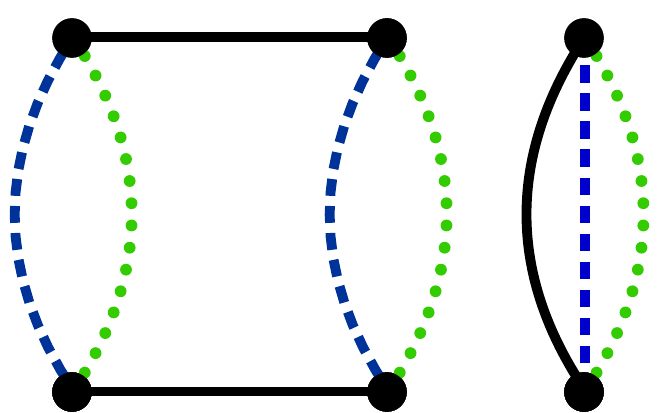}\\
		\end{center}
	\end{minipage} & \begin{minipage}{3cm}
		\begin{center}
			\includegraphics[scale=.3,angle =90]{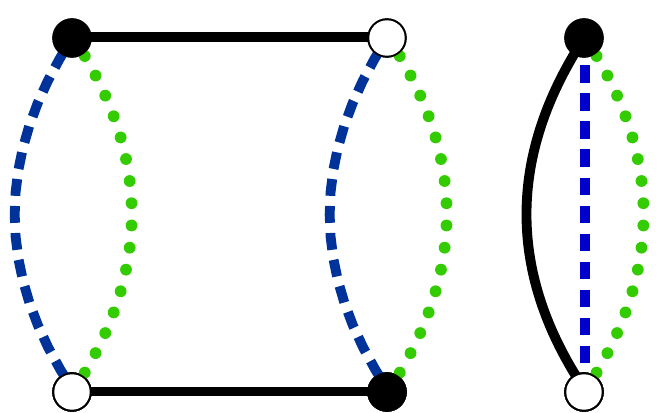}\\
		\end{center}
	\end{minipage}
\\ \hline
6 &  $T_{a_1 a_2 a_3} T_{b_1 a_2 b_3} T_{b_1 b_2 a_3} T_{a_1 b_2 b_3} T_{c_1 c_2 c_3} T_{c_1 c_2 c_3}$ & \begin{minipage}{3cm}
		\begin{center}
			\includegraphics[scale=.3,angle =90]{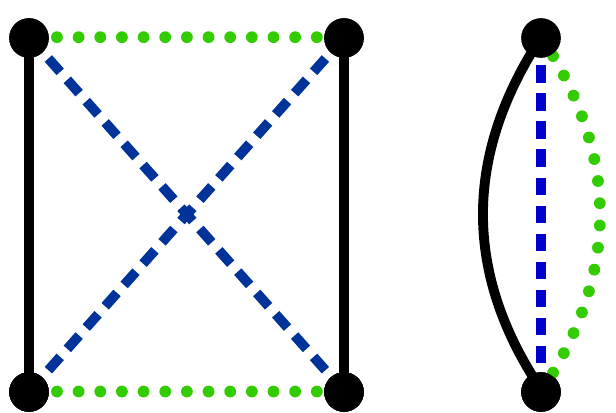}\\
		\end{center}
	\end{minipage} & $-$
\\ \hline
\end{tabular}
\caption{\label{tablecombstruct}  Graphical representation of all invariants allowed by $O (N^\prime)^{\otimes 3}$ ($U (N^\prime)^{\otimes 3}$) symmetry for a rank-3 real (complex) tensor model up to sixth order in the tensors. For the real model, all tensors are represented by black vertices while in the complex model, black and white vertices are used to distinguish the tensor $T$ and its complex conjugate $\bar{T}$. In this case, the algebraic representation of the invariants has to be written as contractions of $T$-tensors with $\bar{T}$-tensors and takes the analogous form to the real expressions provided explicitly.
}
\end{table}

Note that the theory space includes multi-trace interactions. This name derives from the rank-2, i.e., matrix-model case, where interactions take the form ${\rm Tr}\, T_{a_1b_1}...T_{a_nb_n}\cdot...\cdot {\rm Tr}\, T_{a_1b_1}...T_{a_mb_m}$. In the case of higher rank, similarly combinatorially disconnected interactions are part of the theory space. These are generated by the flow, even if they are not included in a truncation. There is no symmetry principle (that we are aware of) that allows to set the corresponding couplings to zero.

\subsection{Regulator \& symmetry breaking}
The key ingredient to set up the flow equation is the regulator, or ``infrared" suppression term. In this context, infrared means low values of indices. Accordingly, the regulator should satisfy the two limits
\begin{itemize}
\item[1)] $R_N(\left\{a_i\right\}) \rightarrow 0$ for $N/\sum^{d}_{i=1}a_i \rightarrow 0$,
\item[2)] $R_N(\left\{a_i\right\}) >0$ for $\sum^{d}_{i=1}a_i/N<1$,
\item[3)] $R_N(\left\{a_i\right\}) \rightarrow \infty$ for $N \rightarrow N' \rightarrow \infty$.
\end{itemize}
The first condition ensures that ``UV" modes are unsuppressed. It also ensures that no modes are suppressed once the IR cutoff scale $N$ is lowered to zero. The second condition enforces that ``IR" modes are suppressed. The third condition ensures that in the limit of infinite cutoff, the effective action essentially reproduces the classical action.  
The three conditions
can be achieved with different so-called shape functions, i.e., different choices of $R_N (\left\{a_i\right\})$. The arguably simplest choice is 
\be
R_N(\left\{a_i\right\})=\left( \frac{N^r}{\sum^{d}_{i=1}a^p_i}-1\right)\theta\left(\frac{N^r}{\sum^{d}_{i=1}a^p_i}-1\right),\label{eq:reg}
\ee
where $r,p>0$. 
While there are optimization criteria for a similar shape function at lowest order in the derivative expansion in the continuum \cite{Litim:2000ci}, it has not yet been investigated what form an optimized cutoff takes for tensor models. \\
As a generalization, one might consider the argument of the regulator to be $N^r/(a^{r_1}+b^{r_2}+c^{r_3})$, which should result in three combinations of the four parameters $r, r_1, r_2, r_3$ to appear in the beta functions. Demanding a discrete symmetry of the indices fixes  
$r_1=r_2=r_3=p$. 

The introduction of the regulator term necessarily breaks the symmetry of the model, as the $O(N')\otimes...\otimes O(N')$ (or $U(N')\otimes...\otimes U(N')$) symmetry requires all index positions to be treated on an equal footing. 
Setting up the RG flow is therefore incompatible with the unbroken symmetry, leading to an enlargement of the theory space. Specifically, the invariants in Eq.~\eqref{eq:invariants}
 are generalized and include
\be
\mathcal{O}^{(n)}_{\rm SB} =f(a_1,...,d_n) T_{a_1 b_1...d_1}...T_{a_n b_n...d_n} \mathcal{C}_{a_1...a_n, b_1...b_n,...,d_1...d_n},\label{eq:invariants2}
\ee
with functions $f(a_1,...,d_n)$ encoding the explicit index dependence. Yet there is an important difference to a setting where the symmetry is broken from the outset, and which features the same theory space. It lies in a modified Ward identity that accounts for the symmetry-breaking introduced by the regulator. It selects a hypersurface in the larger theory space on which the full symmetry is recovered at the IR endpoint of the flow. Although the regulator vanishes in this limit, this is not sufficient to restore the symmetry, since the regulator has introduced symmetry violations in the flow at all finite scales. To compensate these, the initial condition  for the flow, set in the UV needs to break the symmetry in a specific way that is dictated by the Ward-identity. Therefore a fixed point of the RG flow simultaneously needs to solve the modified Ward identity in order to lead to a symmetric IR limit. 
This requirement cannot necessarily be imposed on truncations: While the exact flow equation and Ward-identity are compatible, the Ward-identity in general requires other terms to be present in the truncation than the flow equation provides.

In matrix models, a simple solution of the Ward-identity was discovered \cite{Eichhorn:2014xaa}: As symmetry-breaking is not introduced through tadpole diagrams (i.e., the leading order contributions to the beta functions in an expansion in couplings) in matrix models, the theory space is not enlarged in the tadpole-approximation. Beyond rank 2, such a simple 
solution
is no longer possible, as even the tadpole approximation generates symmetry-breaking terms.

\subsection{Bootstrap strategy for consistent truncations}

To characterize a universality class, at least all non-irrelevant critical exponents must be calculated. Accordingly, the set of all couplings which have a significant overlap with a relevant or marginal direction must be included in a minimal truncation. A priori, this set is not determined at an interacting fixed point. 
In practice, the following strategy is available:
one starts with an assumption about a systematic division of theory space into relevant and irrelevant directions. A reliable truncation should  at least include all couplings which are expected to be relevant as well as the leading irrelevant ones. If the beta functions in this truncation feature a fixed point, the critical exponents at the fixed point indicate whether the initial assumption about relevant couplings holds. If this is the case, terms beyond the truncation are expected to most likely only provide  
 subleading
corrections to the relevant critical exponents.

A particularly useful assumption is that of near-canonical scaling which allows one to use the canonical dimension as a guiding principle. This assumption works very well for a large class of fixed points and implies essentially that low orders in a vertex expansion are sufficient to obtain quantitative estimates of the critical exponents. The underlying reason is that for these cases, the mechanism that induces the fixed point is a balance between canonical scaling and leading-order quantum corrections. This mechanism is at work as soon as one departs from the critical dimension of a particular interaction, and generates a 
UV (IR) attractive fixed point if the coupling is asymptotically free (trivial) in its critical dimension. Examples include Yang-Mills in $d=4+\epsilon$, the Gross-Neveu model in $d=2+\epsilon$ for the former and the Wilson-Fisher fixed point for the latter, see, e.g., \cite{Peskin:1980ay,Gies:2003ic,Morris:2004mg,Wilson:1971dc,Gawedzki:1985ed,Kikukawa:1989fw}.

For the search of a quantum gravity fixed point in the tensor model theory space, the canonical dimension could be a useful guiding principle. The motivation for this comes from the hope that the universality class discovered for quantum gravity in the continuum where metric fluctuations are summed over appears to be near-canonical. To match the corresponding spectrum of scaling exponents, one would expect a near-canonical scaling also on the tensor model side \footnote{ Note that if the continuum limit in tensor models can be taken and provides a well-defined continuum space(time), this is equivalent to a version of quantum gravity being asymptotically safe. We stress that asymptotic safety is a general scenario for path integrals. As such it is not tied to one particular choice of configuration space and might even be realized in several distinct configuration spaces. Therefore the continuum path integral corresponding to tensor models might well be one that includes a summation over (a subset of) topologies and is therefore not the same asymptotically safe gravity model that appears to exist according to continuum studies summing over metric fluctuations only.}.
The continuum asymptotic safety regime has been studied intensively and there is mounting evidence that the non-Gaussian fixed point explored in 
that
approach  
features
near-canonical scaling, the largest anomalous scaling is about $2$  while the difference of quantum to canonical scaling goes to zero for the couplings of $\sqrt{g}R^n$ with $n>3$, see, e.g. \cite{Falls:2013bv,Falls:2014tra,Falls:2018ylp,Eichhorn:2018ydy}. It is therefore  
 a well-motivated starting point
to assume that no operator with, e.g., canonical dimension $-4$ (or slightly more negative) can have significant overlap with a relevant direction at the quantum gravity fixed point in tensor models. This leads to a truncation ansatz in which one includes all operators up to this scaling. One can then search for a fixed point with quantum-gravity characteristics and check explicitly whether the near-canonical scaling assumption is justified. 

Having found such a semi-perturbative fixed point in a truncation, one needs to check whether this fixed point is a truncation artifact, i.e., that the RG-flow at that point simply 
is
parallel to the projection onto the truncation. One can obtain hints about this by (1) varying the regulator andthe way in which one projects onto the truncation ansatz and (2) enlarging the truncation. If a fixed point is stable under variations of the regulator and projection rule and if it appears with the same features in larger truncations, then it is unlikely that the fixed point is a truncation artifact. A larger truncation also allows one to re-check the assumption of near canonical scaling. Ideally one finds that the deviation from canonical scaling decreases for the new operators which suggest that canonical scaling becomes a better and better assumption for the operators not included in the truncation. A similar strategy was successfully applied to the semi-perturbative UV-attractor of the Grosse-Wulkenhaar model \cite{Sfondrini:2010zm}, where  
it was indeed possible
to bound the deviation from canonical scaling. Deriving such a bound  for tensor models would complete the bootstrap approach.

\subsection{In practice: The $\mathcal{P}\mathcal{F}$ expansion}

The FRG equation \eqref{eq:floweqtensors} is an equation for the effective action functional, which involves inverting a field-dependent operator and taking the regulated trace over  the eigenvalues of the field- and index-dependent two-point function
A very useful strategy to perform these two operations is the $\mathcal{P}\mathcal{F}$-expansion, which is a Taylor expansion of the RHS of the flow equation in the tensor $T_{abc}$ around the vanishing field configuration $T_{abc}\equiv 0$.
 To obtain the expansion, 
 we rewrite the regularized
inverse two-point function that enters the flow equation \eqref{eq:floweqtensors} as
\begin{equation}
\Gamma^{(2)}_{N, abcdef} [T] + R_{N,abcdef} = \underbrace{\Gamma^{(2)}_{N, abcdef} [T=0] + R_{N,abcdef}}_{\mathcal{P}} + \underbrace{\Gamma^{(2)}_{N, abcdef} [T] - \Gamma^{(2)}_{N, abcdef} [T=0]}_{\mathcal{F}}\,,
\end{equation}
where we use a shorthand notation $\Gamma^{(2)}_{N, abcdef} = \delta^2\Gamma_N /\delta T_{abc}\delta T_{def}$. Thence, the flow equation \eqref{eq:floweqtensors} is expressed as
\begin{equation}
\partial_t \Gamma_N = \frac{1}{2}\mathrm{Tr}\left[\left(\mathcal{P}+\mathcal{F}\right)^{-1}\partial_t R_N\right] = \frac{1}{2}\mathrm{Tr}\left[\left(\partial_t R_N \right)\mathcal{P}^{-1}\right]+\frac{1}{2}\sum^{\infty}_{n=1}\mathrm{Tr}\left[(-1)^n \left(\partial_t R_N \right)\mathcal{P}^{-1} \left(\mathcal{P}^{-1}\mathcal{F}\right)^n\right]\,,
\end{equation}
where we suppressed the tensor indices for simplicity and expanded the inverse two-point function as a geometric series. This way of writing the RHS of the flow equation is very useful when one considers finite polynomial truncations in $T_{abc}$, because in this case 
one can truncate the sum at finite order.  
All further terms of the sum would possess more tensors than the monomials in the truncation. 

\section{Large $N$ scaling dimensions}\label{sec:Scalingdim}

In settings with a background, where the RG flow corresponds to a local coarse graining, one RG step is literally a scale transformation. Accordingly, the canonical scaling dimensions of couplings are their mass dimensions. These can be determined prior to studying the actual RG flow. In the background independent setting, there is no notion of locality or spacetime and accordingly all couplings 
are dimensionless in terms of units of length or mass, and no notion of mass dimension exists.
Yet, mass dimension is not the notion of dimensionality that is relevant to a pregeometric RG flow anyways. Instead, it is a consistent scaling with $N$ that is central here. This scaling is not determined a priori. Nevertheless, one can determine it in two steps:
\begin{enumerate}
    \item Since the purpose of the FRG setup is the investigation of
the large $N$-behavior of the tensor model, we need to scale the
coupling constants in such a way that the beta functions admit a $1/N$
expansion. This gives a stack of coupled inequalities, which exclude
most scaling prescriptions. Imposing the additional requirement that no interactions should be artificially decoupled from the system uniquely fixes all but one scaling dimensions.
    \item A further condition comes from the geometric interpretation of tensor models. Specifically, the interpretation in terms of the 
Regge-action of the triangulation that is associated to each tensor-model Feynman graph  is only possible for a particular scaling of
the associated coupling constant with $N$.
\end{enumerate}
We will now present these two steps in more detail and determine the
scaling dimension for the tensor models of quantum gravity.

For the first step, let us briefly 
return
to the background dependent continuum setting. There, 
the flow equation automatically provides a scaling dimension. It arises by demanding that the beta functions form an autonomous system, such that, after an appropriate rescaling of the couplings, the explicit dependence on the scale drops out. As a specific example, consider the beta function for the Newton coupling $\bar{G}$, which reads
\be
\beta_{\bar{G}} = \# k^{d-2}\bar{G}^2,
\ee
to leading order in $\bar{G}$ with $\#<0$ , \cite{Hawking:1979ig,Gastmans:1977ad,Christensen:1978sc,Kawai:1989yh,Codello:2008vh,Falls:2015qga}. Demanding independence from $k$ provides the scaling dimension and is in agreement with the mass dimensionality. The dimensionless coupling takes the form
\be
G = \bar{G}k^{d-2}.
\ee
Without knowing anything about mass dimensionality, one can thus alternatively fix the canonical scaling dimensions of couplings by demanding that the beta functions form an autonomous system. This strategy is applicable to the pregeometric setting. For instance, the coupling $\bar{g}_{4,1}^{2,1}$ of the interaction $T_{abc}T_{ade}T_{fde}T_{fbc}$ in a real rank 3 model has the beta function 
\be
\beta_{\bar{g}_{4,1}^{2,1}} = \# N^\frac{2r}{p} \left(\bar{g}_{4,1}^{2,1}\right)^2+ \mathcal{O}(N^0),
\ee
resulting in
\be
g_{4,1}^{2,1}= N^\frac{2r}{p} \bar{g}_{4,1}^{2,1}.\label{eq:scaling4121}
\ee
Note that fixing the scaling dimensions in this way is 
possible in the large $N$ limit, but not at finite $N$. This is a consequence of the fact that at any given order in the couplings, different orders in $N$ appear. As we explicitly use the large $N$-limit, this only results in an \emph{upper bound} on the scaling dimensions. Choosing scaling dimensions below this upper bound also results in autonomous beta functions in the large $N$ limit\footnote{For some couplings, however, the set of inequalities to be fulfilled for a well-defined large-$N$ limit also provide lower bounds. 
A consistent assignment of scaling dimension for a coupling therefore generically requires the inspection of several beta functions which depend on this coupling.
}. Yet, for this choice the corresponding interactions decouple from the beta functions. The ``most interacting" system, where no interactions are suppressed artificially, is achieved when the scaling dimensions are chosen as the upper bounds. 

As is evident from Eq.~\eqref{eq:scaling4121}, the thus-determined scaling dimensions depend on the parameters $r$ and $p$ in Eq.~\eqref{eq:reg}. 
Insight into the physics allows to fix the ratio $r/p$. For instance, for the geometric interpretation of tensor models, the interpretation of the dual picture in terms of dynamical triangulations results in a relation of the couplings of the tensor model and the scale $N$ to the couplings of the Regge action. This relation only works for a specific choice of canonical scaling for the leading coupling (i.e., one of the quartic couplings). In turn, this scaling dimension fixes $r/p$. It turns out that this is $r/p=1$ for a rank $d$ tensor model, if $d$ is the dimension entering the corresponding Regge action in the continuum picture. Yet, for those fixed points in the tensor model that show dimensional reduction to a matrix model\footnote{ See Sect.~\ref{sec:rank3} and \ref{sec:rank4}.},  
$r/p<1$ is the correct choice.

As a specific example for how the geometric interpretation fixes $r/p$, consider
the
possibly simplest quantum-gravity tensor model,  the so-called rank 3
colored complex model \cite{Gurau:2011xp} defined through the action
\begin{equation}
 S(T,\bar T)=\sum^3_{i=0}T^i_{abc}\bar{T}^i_{abc}
   +N^{-3/2}\left(\lambda\,
T^0_{abc}T^1_{ade}T^2_{fbe}T^3_{fdc}+c.c.\right).
\end{equation}
The Feynman-diagram expansion of this model yields the amplitude
\begin{equation}
 A(\gamma)=N^{N_{1}-3/2N_3}\,(\lambda\bar\lambda)^{N_3/2},
\end{equation}
where $N_3$ denotes the number of 3-cells\footnote{An $n$-cell of a triangulation is an $n$-dimensional simplex that appears as an elementary building block of the triangulation of a $d$-dimensional pseudo-manifold. For instance, a 3-cell is a tetrahedron, a 2-cell a triangle, which appears in the boundary of a tetrahedron, a 1-cell an edge, which appears in the boundary of a triangle, and so on.} in the triangulation
$\Delta(\gamma)$ associated with the colored Feynman graph $\gamma$ and
where $N_1$ denotes the number of 1-cells. Comparing this with the Regge
action $S_R[\Delta]=\kappa_3 N_3-\kappa_1N_1$ of a triangulation
$\Delta$ allows us to identify the coupling constants as
$\kappa_1=\ln(N)$ and $\kappa_3=\frac 3 2\ln(N)-\frac 1 2
\ln(\lambda\bar\lambda)$. The uncolored models that we investigate with
the FRG
are obtained by integrating out all but the last color. The
scaling $N^{-3/2}$ of the coupling constant $\lambda$ then implies the
scaling $N^{-2}$ for the cyclic melonic interactions. In other
words the geometric compatibility condition fixes
$r/p=1$.

\section{Benchmarking the FRG in matrix models}\label{sec:matrixmodel}
In two-dimensional quantum gravity, the relevant critical exponent of the double-scaling limit is known. In this limit, the continuum limit in dynamical triangulations can be taken in such a way that all topologies contribute. For reviews and introductions, see, e.g., \cite{DiFrancesco:1993cyw,ZinnJustin:2002ru,Ginsparg:1991bi,Ambjorn:1994yv,Marino:2004eq}. The matrix model that is dual to dynamical triangulations can be chosen to be  Hermitian $N \times N$ matrices $\varphi$, with the generating functional given by
\be
Z= \int \mathcal{D}\varphi\, e^{N \left( -\frac{1}{2}{\rm Tr}\varphi^2+ \frac{g_4}{4}{\rm Tr}\varphi^4\right)}.
\ee
The double-scaling limit requires taking $N \rightarrow \infty$, while holding
\be
\left(g_4-g_{4\, \rm crit}\right)^{\frac{5}{4}}N = \rm const,
\ee
where $g_{4\,\rm crit}$ is the critical value of the coupling.
This can be rewritten in the form
\be
g_4(N) = g_{4\, \rm crit} + c\, N^{-\frac{4}{5}}.
\ee 
This is structurally similar to the leading-order scaling of couplings in the vicinity of a fixed point of the RG flow. Accordingly, one is led to
identify
\be
\theta=\frac{4}{5},
\ee
as a relevant critical exponent. This similarity prompted the authors of  \cite{Brezin:1992yc} to set up a pregeometric RG flow in matrix size $N$. In that paper as well as the follow-up works \cite{Alfaro:1992nq,Ayala:1993fj,Higuchi:1993pu,Higuchi:1994rv,Higuchi:1993nq,Dasgupta:2003kk}, the coarse-graining was implemented explicitly by integrating out the outermost rows and columns of the matrices in a Gaussian approximation.

In \cite{Eichhorn:2013isa}, the flow equation \eqref{eq:floweqtensors} in the pregeometric setting was first derived. Applying it to truncations of a single-trace form, $\Gamma_N = \sum_{i=2} g_{2i} {\rm Tr}\phi^{2i}$, yielded a critical exponent that approaches $\theta=1$ from above. Extending the truncation to multitrace operators does not improve the estimate, but instead makes it worse. The critical exponent $\theta=0.8$ for gravity is first reproduced at the first multicritical point  \cite{Eichhorn:2014xaa}, which corresponds to gravity coupled to conformal matter \cite{Kazakov:1989bc}. 

Instead of reviewing these results in greater detail, here we explore an alternative prescription to calculate the critical exponents that leads to a significant improvement in the estimate. This prescription was already explored in \cite{Eichhorn:2017xhy} for tensor models, and has been put forward for continuum QFTs  in \cite{Boettcher:2015pja}. 
It consists in keeping the anomalous dimension 
$\eta = -\, N\partial_N\, {\rm ln}Z_N$ constant while calculating the stability matrix, i.e.,
\be
\tilde{\theta}_I = -{\rm eig} \left(\left(\frac{\partial \beta_{g_i}}{\partial g_j}\right)_{\eta} \right)\Big|_{\vec{g}=\vec{g}_{\ast}}.\label{eq:newtheta}
\ee
The notation $()_{\eta}$ indicates that the derivative is taken at fixed $\eta$.
The alternative, more standard prescription differs by including derivatives of $\eta$ and will be denoted by $\theta_I$ to clearly differentiate between the two. As a specific example, consider the case where a coupling $g_i$ already corresponds to an eigendirection at a fixed point. No off-diagonal elements of the stability matrix contribute to its critical exponent, such that
\bea
\tilde{\theta}&=& - \left(\left(\frac{\partial \beta_{g_i}}{\partial g_j}\right)_{\eta} \right)\Big|_{\vec{g}=\vec{g}_{\ast}},\\
\theta&=&-\left(\frac{\partial \beta_{g_i}}{\partial g_i}\right)\Big|_{\vec{g}=\vec{g}_{\ast}} = \tilde{\theta} +\left(\frac{\partial \beta_{g_i}}{\partial \eta} \frac{\partial \eta}{\partial g_i}\right)\Big|_{\vec{g}=\vec{g}_{\ast}}.
\eea

In \cite{Boettcher:2015pja} it was observed that a scaling relation for critical exponents in the $O(N)\oplus O(M)$ model, which is known to hold for the epsilon-expansion  \cite{Wilson:1971dc,LeGuillou:1977rjt,Guida:1998bx}, is only satisfied for the FRG in truncations of the full flow to the local potential approximation plus anomalous dimension for the prescription in Eq.~\eqref{eq:newtheta}. The more standard prescription leads to small violations of the scaling relation in those truncations.

\begin{figure}[!t]
\centering
\includegraphics[width=0.5\linewidth]{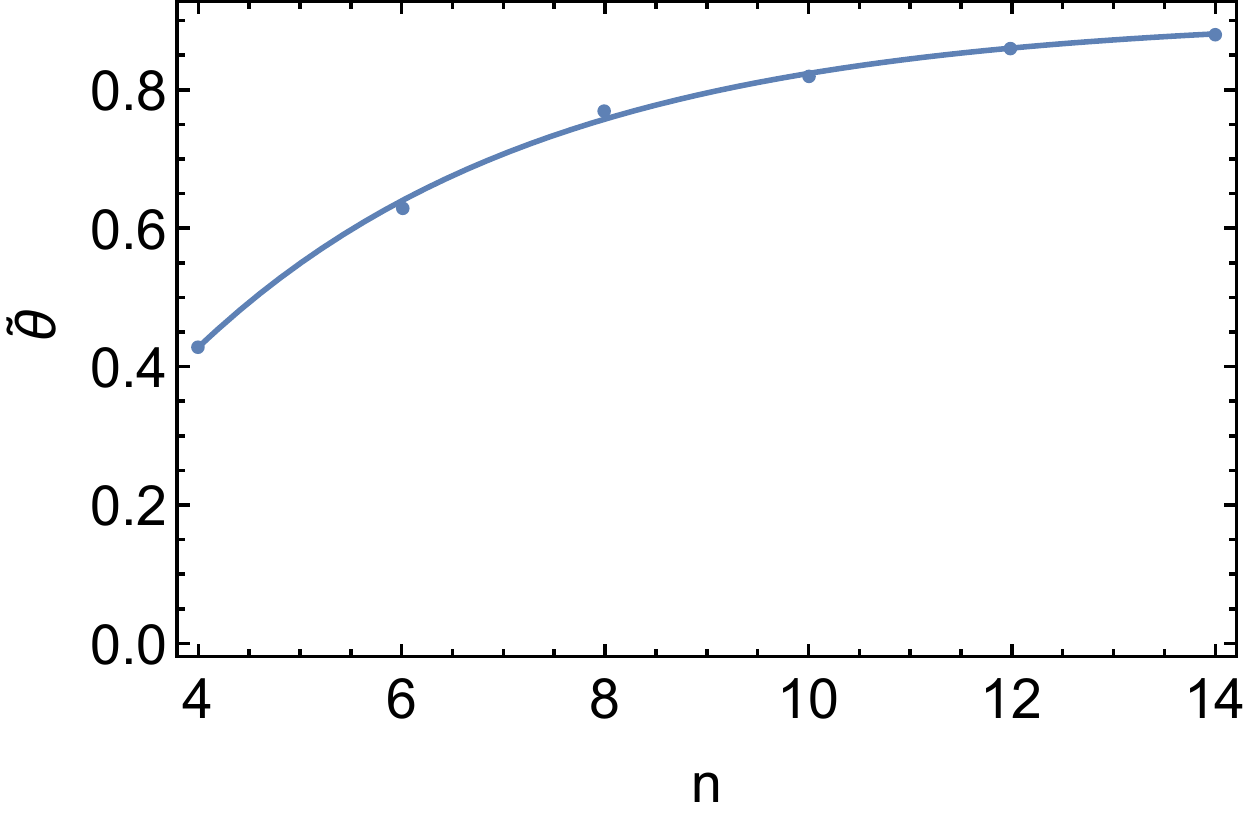}
\caption{\label{fig:critexpMM}The relevant critical exponent in the matrix model according to the prescription Eq.~\eqref{eq:newtheta} as a function of truncation order in a truncation of the form $\Gamma_N = \sum_{i=1}^ng_{2i}\, {\rm Tr}\phi^{2i}$. }
\end{figure}

Here, we show that the $\tilde{\theta}$-prescription gives improved results for the critical exponent of the double-scaling limit, resulting in only 14 \% deviation already in a calculationally very straightforward truncation, cf.~Fig.~\ref{fig:critexpMM}.

The results for the critical exponent as a function of truncation order in Fig.~\ref{fig:critexpMM} appear to be fit well by a function of the form 
\be
\tilde{\theta}(n) = a- b\, e^{-c\, n},
\ee
with fit parameters $a=0.91$, $b=1.54$ and $c=0.29$. An extrapolation to $n \rightarrow \infty$, which is the complete single-trace subsector of theory space, yields $\tilde{\theta}(n\rightarrow \infty)=0.91$, which is only a 14 \% deviation from the exact result $\theta=0.8$. Whether this is accidental, or whether there is a deeper reason why the $\tilde{\theta}$ prescription works better for matrix and potentially also tensor models, remains to be explored in the future. 

One source of systematic errors for the critical exponent is  the breaking of the $U(N)$ symmetry of the matrix model through the regulator \cite{Eichhorn:2014xaa}. This can be seen by the fact that the $U(N)$-Ward-identity obtains a non-vanishing RHS through the introduction of the regulator: 
\be
\mathcal G_\epsilon \Gamma_N=\epsilon\, \textrm{Tr}\left(\frac{[A,R_N]}{\Gamma^{(2)}_N+R_N}\right),
\ee
 where $A$ is the generating matrix of an infinitesimal unitary transformation, which generates the transformation $\mathcal G_\epsilon$. By generating we mean that a unitary transformation $U=\exp(i\,\epsilon\,A)$ transforms the matrix $\phi$ as $\phi \mapsto U^\dagger.\phi.U = \phi+i\,\epsilon\,[A,\phi]+\mathcal O(\epsilon^2)$. This implies that the RG flow generates symmetry breaking operators even if the initial condition is a $U(N)$ symmetric action. In particular the relevant directions will acquire contamination by these symmetry breaking operators. Hence, when  investigating  the large $N$-limit with the FRG, one has to include these symmetry-breaking operators to find accurate critical exponents. 

Including the symmetry breaking operators into a truncation and distinguishing them from the symmetric operators by a projection on the truncation is a technically rather challenging task. Fortunately, there is a self-consistent work-around in the case of matrix models that gives surprisingly stable results \cite{Eichhorn:2014xaa}: It is based on the observation that tadpole diagrams of $U(N)$-symmetric operators do not generate symmetry-breaking operators for the rank-2-case. In other words, the tadpole approximation to the broken $U(N)$-Ward-identity is solved by a symmetric effective average action. Using the tadpole approximation in a single trace truncation allows one to find the infinite series of so-called multicritical points. The $m$-th multicritical point is a fixed point with $m$ non-vanishing couplings at the fixed point whose fixed-point values occur with alternating sings and whose critical exponents are $\theta^{(m)}_n=\frac{n}{m}$. We see that the largest critical exponent (the pure gravity exponent) is still $\theta^{(m)}_m=1$ in the single trace truncation. However, including multitrace operators in the truncation yields improved results for the largest critical exponents $\theta^{(m)}_m=0.80...0.82$ at the $m=2,3,4,...$ multicritical fixed points. 
 For this leading critical exponent, one therefore obtains an estimate that deviates from the exact value by only 3\%, which is a rather high precision.  The subleading relevant critical exponents at the multicritical points are not reproduced with comparable precision in this truncation. Nevertheless, we interpret the precision of the leading relevant exponent as a signature that the FRG can successfully pass the benchmark test posed by rank 2 models.
Only at the double scaling limit, i.e., the $m=1$ fixed point, one obtains $\theta^{(1]}_1=1$. This relatively large discrepancy of the critical exponents from $0.8$ can be explained by the fact that the tadpole approximation does only capture effects from the tree-level truncation, which contains only one coupling. Given such a small truncation, it is actually remarkable to obtain the values of the critical exponent with $25\%$ accuracy.

As a consequence of universality, specific fixed points in tensor models can also reproduce the matrix-model results. This is due to the fact that the shape of the building blocks is not relevant for the continuum limit. Therefore even higher-dimensional building blocks can reproduce a lower-dimensional continuum limit, at a point in theory space where the effective dynamics ``flattens" these building blocks in an appropriate way. To recover the lower-dimensional scaling, the canonical scaling dimensions of the model have to be adjusted by choosing $r/p<1$ in Eq.~\eqref{eq:reg}.
In that choice, and for the prescription Eq.~\eqref{eq:newtheta}, the matrix-model exponent is approximately recovered from the fixed points in tensor models, see Sect.~\ref{sec:rank3} and \ref{sec:rank4}.

\section{Charting three dimensions from a tensor-model point of view}\label{sec:rank3}

An important motivation for RG studies of the large-$N$- behavior of tensor models is the search for a continuum limit that can be associated with quantum gravity. The first step in the systematic program that can lead to the confirmation or refutation of the conjecture that there might exists a continuum limit in tensor models which corresponds to quantum gravity, is a systematic investigation of theory spaces. Varying the number of tensor fields, the rank of the tensors and symmetry-structures provides a number of different theory spaces which one can then investigate with the FRG. The first step in the FRG investigation of a theory space consists of finding tentative candidates for universal fixed points. This provides insight into which interaction structures could be of particular importance for a continuum limit. Below, we discuss the status of this systematic program in more detail for rank-3 tensor models.

In summary, by investigating a complex uncolored model, i.e., a model with $U(N')\otimes U(N')\otimes U(N')$ symmetry, and a real uncolored model, i.e., a symmetry group of the form $O(N')\otimes O(N')\otimes O(N')$, we discover that certain classes of fixed points are shared. In particular, we find fixed points that exhibit a form of dimensional reduction and evidence that these fixed points are not truncation artifacts. Crucially, the real model features a new, tetrahedral interaction, cf.~the third entry in Tab.~1, introduced by Carrozza and Tanasa  \cite{Carrozza:2015adg}, and later taken up in \cite{Klebanov:2016xxf} for an SYK-type model. This interaction appears to be key for the generation of a fixed point which does not appear to feature dimensional reduction and therefore constitutes a tentative candidate for a continuum limit for three-dimensional quantum gravity. We stress that of course it requires much more than just the discovery of the fixed point to establish its relevance for three-dimensional quantum gravity; finding a fixed point without dimensional reduction is a necessary but not sufficient step in linking tensor models to a well-behaved phase of quantum gravity.

\subsection{Dimensional reduction in tensor models}

The absence of a background geometry permits that tensor models exhibit phenomena that do not appear in local quantum fields theories. The first of these is the dynamical generation of multi-trace operators, which correspond to tensor model vertices with a geometric interpretation as boundaries formed by disconnected pieces of geometry (such as, e.g., the two circles in the boundary of a cylinder). These multitrace operators are however generated by connected Feynman diagrams. 
For instance, a matrix connected matrix model Feynman diagram may be dual to the triangulation of a cylinder connecting the two circles in the boundary. The corresponding interactions 
are thus generically generated by the flow, and are part of 
the quantum effective action. In particular one finds disconnected tensor invariants with $2n$ tensors of the form $(T_{abc}T_{abc})^n$, which possess an enhanced  $O(N'^3)$-symmetry, reducing the tensor model to a vector model and producing non-Gaussian fixed points, which do not represent extended three-dimensional geometries. 

In \cite{Eichhorn:2018ylk} we identified a mechanism  that can be realized at suitable fixed points and prevents the production of multi-trace operators. It is based on the observation that the generation of $(T_{abc}T_{abc})^2$ from connected vertices requires two cyclic 4-melons with distinct preferred colors to be nonzero. Thus, the fixed points in
the theory space with the enhanced $O(N')\otimes O(N'^2)$ symmetry exhibited by the cyclic melons with one preferred color do not possess  non-vanishing multi-trace operators. However, this theory space 
exhibits dimensional reduction. Dynamical dimensional reduction at high energies is an intriguing phenomenon in several models of quantum gravity, see, e.g., \cite{Carlip:2017eud}, that are four-dimensional at large scales. In tensor models, dimensional reduction differs in that it appears to be realized at certain classes of fixed points in rank-3 and  rank-4 models, such that the continuum limit is not a candidate for three- or four-dimensional quantum gravity. (Although not yet explored explicitly, the same result should be true in any rank $d>2$.) Specifically, an enhancement of the $U(N')\otimes ...\otimes U(N')$ (or $O(N')\otimes ...\otimes O(N')$) symmetry to an  $U({N'}^2)\otimes...\otimes U(N')$ ($O({N'}^2)\otimes...\otimes O(N')$) symmetry goes hand in hand with an effective ``fusion" of two indices into one ``superindex", such that the model effectively reduces to a matrix model. This occurs at fixed points at which only cyclic melons (single trace, multi-trace or both) of one preferred color are present. Because of the enhanced symmetry, it is always consistent to set all other interactions to zero, as one can also check by inspecting the beta functions. To fully establish the dimensional reduction, the critical exponents of the matrix model should also be reproduced. Here, the freedom in choosing $r$ and $p$ in Eq.~\eqref{eq:reg} becomes crucial: A matrix model features different canonical dimensions than a tensor model, essentially due to the reduced rank. The canonical dimensions are functions of $r/p$. To probe the matrix-model limit of rank-3 tensor models, one should choose $r/p=1/2$ to obtain the canonical dimensions appropriate for a matrix model. With this choice for the scaling of the regulator, and for the prescription Eq.~\eqref{eq:newtheta}, the matrix-model exponent is approximately recovered from the fixed points in tensor models. 

In particular, the fixed points  reported in
Tables~\ref{cycmelsttable2}, \ref{multitracebubbletable2} and \ref{tab:cmmmtFP} feature dimension reduction. They were obtained for the rank-3 real model. For those which appear both in the complex and in the real model, both sets of values are shown.
The tables display fixed points as well as critical exponents values obtained in the hexic truncation of the rank-3 real model. In this truncation, all $O(N^\prime)\otimes O(N^\prime)\otimes O(N^\prime)$ symmetric interactions are included in the effective average action. The resulting truncation has 21 couplings. 
Further details can be found in \cite{Eichhorn:2017xhy,Eichhorn:2018ylk}.

\begin{table}[h!]
		\begin{tabular}{|c|c|c|c|c|c|c|c|c|c|c|c|c|c|c|c| } 
			\hline
			 ${g_{4,1}^{0}}^*$ & ${g_{4,1}^{2,1}}^*$ & ${g_{4,1}^{2,2}}^*$ & ${g_{4,1}^{2,3}}^*$& ${g_{4,2}^{2}}^*$& ${g^{0,np}_{6,1}}^*$ & ${g^{0,p}_{6,1}}^*$ & ${g^{1,i}_{6,1}}^*$ & ${g^{2,i}_{6,1}}^*$  & ${g^{3,1}_{6,1}}^*$ & ${g^{3,2}_{6,1}}^*$ & ${g^{3,3}_{6,1}}^*$& ${g^{1}_{6,2}}^*$ & ${g^{3,i}_{6,2}}^*$  & ${g^{3}_{6,3}}^*$   \\ \hline 
			 0 & -0.28 & 0 & 0 & 0 & 0 & 0 & 0 & 0  & -0.15 & 0 & 0 & 0  & 0 & 0   \\ 
			\hline
			\hline
 $\theta_1$ & $\theta_2$ & $\theta_3$ & $\theta_{4,5}$  & $\theta_6$ & $\theta_{7,8}$ & $\theta_9$ & $\theta_{10}$ & $\theta_{11,12}$  & $\theta_{13,14}$  & $\theta_{15,16,17}$ & $\theta_{18}$ & $\theta_{19}$ & $\theta_{20,21}$  & $\eta$\\ \hline 
			 2.19 & -0.03 &  -0.69 &  -1.19  & -1.68 & -1.78 & -2.08 & -2.15 & -2.18 & -2.28 & -2.78  & -2.94 & -2.98 & -3.18 &-0.41 \\  
			\hline
		\end{tabular}\\
		\caption{This fixed point only features cyclic melonic couplings and has one relevant direction. It features dimensional reduction to a matrix model. For the complex model, the first four critical exponents are $\theta_1=2.14$, $\theta_2=-0.59$, $\theta_{3,4}=-1.26$. The slight numerical difference to the real case is due to a different choice of projection scheme. The second critical exponent of the real model misses from the universality class obtained from the complex model and is to be attributed to the additional presence of the interaction associated with $g_{4,1}^0$. Choosing $r/p=1/2$ leads to $\theta=1.09$ for the leading critical exponent. Using the prescription described in Sec.~\ref{sec:matrixmodel} the value is $\tilde{\theta}=0.63$.}
	\label{cycmelsttable2}
	\end{table}

	\begin{table}[h!]
		\begin{tabular}{ |c|c|c|c|c|c|c|c|c|c|c|c|} 
			\hline
			 ${g_{4,1}^{0}}^*$ & ${g_{4,1}^{2,i}}^*$ & ${g_{4,2}^{2}}^*$& ${g^{0,np}_{6,1}}^*$ & ${g^{0,p}_{6,1}}^*$ & ${g^{1,i}_{6,1}}^*$  & ${g^{2,i}_{6,1}}^*$   & ${g^{3,i}_{6,1}}^*$  & ${g^{3,i}_{6,2}}^*$ & ${g^{1}_{6,2}}^*$  & ${g^{3}_{6,3}}^*$   \\ \hline 
			 0 & 0 & -1.05 & 0  & 0 & 0 & 0 & 0& 0    & 0 & -2.27   \\ 
			\hline
		\end{tabular}\\
		\begin{tabular}{ |c|c|c|c|c|c|c|c|c|c|c| }
			\hline
			 $\theta_1$ & $\theta_2$ & $\theta_{3,4,5}$ &  $\theta_{6,7}$  & $\theta_{8,9,10}$  & $\theta_{11}$ & $\theta_{12,13,14,15,16,17}$ & $\theta_{18,19,20}$  & $\theta_{21}$& $\eta$\\ \hline 
			 3.32 & -0.33 & -0.83 &  -1.24  & -1.74 & -1.82 & -2.24 & -2.32  & -3.28&-0.59 \\ 
			\hline
		\end{tabular}\\
		\caption{
			Only bubble-multi-trace interactions are present in this fixed point, i.e., interactions of the form $(T_{abc}T_{abc})^n$. It has one relevant direction and features dimensional reduction to a vector model. For the complex case, the first four critical exponents read $\theta_1=3.33$, $\theta_{2,3,4}=-0.83$. These were not provided in \cite{Eichhorn:2017xhy}, but can easily be extracted from the beta functions reported in that work.}
	\label{multitracebubbletable2}
	\end{table}

	\begin{table}[h!]
		\begin{tabular}{ |c|c|c|c|c|c|c|c|c|c|c|c|c|c|c|} 
			\hline
			 ${g_{4,1}^{0}}^*$ & ${g_{4,1}^{2,1}}^*$  & ${g_{4,1}^{2,(2,3)}}^*$  & ${g_{4,2}^{2}}^*$& ${g^{0,np}_{6,1}}^*$ & ${g^{0,p}_{6,1}}^*$ & ${g^{1,i}_{6,1}}^*$  & ${g^{2,i}_{6,1}}^*$ & ${g^{3,1}_{6,1}}^*$  & ${g^{3,(2,3)}_{6,1}}^*$ & ${g^{3,1}_{6,2}}^*$  & ${g^{3,(2,3)}_{6,2}}^*$ & ${g^{1}_{6,2}}^*$  & ${g^{3}_{6,3}}^*$   \\ \hline 
		   0 & -0.27 & 0 & -0.05  & 0 & 0  & 0 & 0 & -0.13 & 0 &  -0.05 & 0 & 0 & -0.01   \\ 
			\hline
		\end{tabular}\\
		\begin{tabular}{ |c|c|c|c|c|c|c|c|c|c|c|c|c|c|c|c| }
			\hline
			 $\theta_1$ & $\theta_2$ & $\theta_{3}$ &  $\theta_{4,5}$  & $\theta_{6}$  & $\theta_{7,8}$ & $\theta_{9}$ & $\theta_{10}$  & $\theta_{11,12}$ & $\theta_{13,14}$ & $\theta_{15}$ & $\theta_{16,17,18}$ &$\theta_{19}$ & $\theta_{20,21}$ & $\eta$\\ \hline 
			 2.23 & 0.03 & -0.66 &  -1.16  & -1.66 & -1.74 & -2.04 & -2.12  & -2.16 & -2.24 & -2.73 & -2.74 & -3.10 & -3.12 & -0.42\\  
			\hline
		\end{tabular}\\
		\caption{\label{tab:cmmmtFP}  This fixed point has non-vanishing melonic as well as multi-trace interactions. It features two relevant directions, the second of which has a very small critical exponent. The systematic error of the truncation is expected to be significantly larger than the deviation of $\theta_2$ from zero. The fixed point exhibits dimensional reduction to a matrix model. For the complex case, the first four critical exponents read $\theta_1=2.56$, $\theta_2=0.44$, $\theta_{3,4}=-0.97$.The difference is to be attributed to a difference in projection scheme. Additionally, the real model features an extra critical exponent $\theta_3=-0.66$ due to the presence of the additional interaction $g_{4,1}^0$.}
	\end{table}

The multi-trace operators $(T_{abc}T_{abc})^n$ are invariant under the enhanced  $O(N'^2)\otimes O(N^\prime)$ symmetry. Thus one expects that there exist fixed points at which cyclic melons and and multi-trace operators take non-vanishing fixed point values. This turns out to be true and, indeed in the quartic and hexic truncations, one finds a non-Gaussian fixed point with  $O(N'^2)\otimes O(N^\prime)$ symmetry and non-vanishing fixed-point values for the cyclic melons and multi-trace operators. This fixed point appears to possess two relevant directions, see Table~\ref{tab:cmmmtFP}, whereas the purely cyclic melonic non-Gaussian fixed point only features one positive critical exponent in this truncation
see Table~\ref{cycmelsttable2}. It should be stressed that the deviation of $\theta_2$ from zero at the fixed point in Table~\ref{tab:cmmmtFP} is smaller than the presumed systematic error of the truncation.

As we have introduced the colors it is consistent
to switch off the multitrace interactions for the matrix-model limit. In matrix-model RG flows this has not been possible, as multitrace interactions in matrix models are automatically generated from single-trace ones. Physically, this might suggest that configurations with disconnected boundaries do not have a significant impact on the path integral in two dimensions, as it appears to be possible to reach the same continuum limit both with an without the presence of multitrace interactions.

Note that due to the symmetry breaking induced by the regulator, even at the single-trace cyclic melonic fixed point, interactions with nontrivial index-dependence outside this theory space are generated and could take finite values at the fixed point. The fixed point values of these operators are constrained by the modified
 $O(N^\prime)^{\otimes 3}$-Ward-identity, where the only symmetry breaking term is due to the regulator. This regulator term vanishes in the IR-limit, which implies that the  $O(N^\prime)^{\otimes 3}$-Ward-identity turns into the constraint that all of these index-dependent interactions vanish. 

The analogous argument applies to all other fixed points: These fixed points will exhibit non-vanishing couplings for index-dependent vertices, but their values are constrained by the modified
Ward-identity. In the IR, it turns into the constraint that all couplings associated with index-dependent operators vanish. To reach this point, the initial condition for the RG flow has to be chosen with an appropriate ``amount" of symmetry-breaking operators, such that during the flow, the symmetry-breaking effect of the regulator compensates with that coming from the initial condition.

\subsection{Candidates with potential relevance for three-dimensional quantum gravity}

The fixed points with enhanced $O(N'^2)\otimes O(N')$ and $O(N'^3)$ symmetries appear to exhibit dimensional reduction. 
This might possibly
be compatible with dynamical dimensional reduction in the physical UV limit, i.e., \emph{after} the continuum limit has already been taken if these fixed points would possess a relevant direction that ``inflates" additional dimensions in the IR. However, we consider this possibility unlikely, and consider it more likely that the continuum limit leads to the same topological dimension as the IR-limit of the corresponding spacetime exhibits. Note that the dimensional reduction in the spectral dimension observed in many quantum-gravity approaches is different, and does not imply that there is a reduction in the topological dimension.\\
A different possibility to search for quantum gravity candidate fixed points is to search for fixed points that do not possess such an
enhanced symmetry. In \cite{Eichhorn:2018ylk} we found two possible candidates. These fixed points are isocolored, i.e., they exhibit a global symmetry under color permutation. Note that such a symmetry is not linked to dimensional reduction. In fact, the presence of cyclic melonic interactions with all three different prefered colors is exactly what prevents the merging of two indices to one ``superindex" linked to $O(N'^2)\otimes O(N')$ symmetry.
Another hint about the ``geometricity" of a fixed point might
be the presence of the tetrahedral interaction $T_{abc}T_{ade}T_{fdc}T_{fbe}$. An isocolored fixed point at which this tetrahedral interaction takes a non-vanishing fixed point value may describe the continuum limit of a geometric model. We stress that this
 is not
  sufficient for
  such a fixed point to
  be associated with quantum gravity. The identification as a quantum-gravity candidate can only be made when 
order parameters indicate a geometric interpretation. The fixed point possesses the positive
  critical exponents 
  \be
  \theta_{\pm}=1.35\pm 1.56i\,...\,1.95\pm 0.69 i,\quad \theta_3=0.38\,...\,0.13.
  \ee
   in the full hexic trunction, where the range comes from several different schemes regarding the treatment of the anomalous dimension. We stress that it should not be taken as a complete estimate of the systematic truncation error. The leading critical exponents are 
   roughly compatible with the critical exponents found for the Einstein-Hilbert truncation in three dimensions \cite{Biemans:2016rvp} with $\theta_1\approx 2.5$ and $\theta_2\approx 0.8$ which are also expected to come with significant systematic errors. We caution that this comparison is subject to systematic errors on both sides. In fact, in three-dimensional continuum gravity, fixed-point searches have only been conducted in the Einstein-Hilbert truncation, not including higher-derivative operators. Therefore it is not yet established whether there are indeed only two relevant directions, although the fact that four-derivative curvature invariants are canonically irrelevant could support such a conjecture. Accordingly, the comparison of critical exponents we perform here is to be understood as a proposal for a comparison that will become more meaningful in the future, when systematic errors are significantly reduced on both sides. Here, we only note that within the significant systematic errors that we expect these results to have, the critical exponents of the continuum and the tensor model setting do not appear to be incompatible. 

A second isocolored melonic fixed point with vanishing fixed-point value for the tetrahedral interaction was also found and discussed in the appendix of \cite{Eichhorn:2018ylk}, but with slightly complex values for the coupling constants. The imaginary parts of the fixed point values of the couplings exhibit a scheme dependence that is consistent with vanishing imaginary parts of the couplings, which would make the fixed-point action real and thus physically admissible.

\section{First steps towards background independent four-dimensional quantum gravity}\label{sec:rank4}

In this subsection, we discuss the first results obtained for rank-4 tensor models using the FRG. The purpose of our presentation is illustrative and for this reason, we restrict the analysis to a simple truncation for the effective average action. An extensive analysis employing more sophisticated truncations will be presented elsewhere. 

Studying rank-4 tensor models, whose Feynman diagrams can be identified with 4 dimensional triangulations, is certainly of great importance
from a quantum-gravity perspective. If a suitable continuum limit can be found, they could be candidates for a 
description of the microscopic structure of four-dimensional quantum spacetime. 
While results in tensor models point towards the existence of a branched-polymer phase \cite{Gurau:2013cbh}, Monte Carlo simulations indicate that causal dynamical triangulations could also give rise to extended four-dimensional geometries \cite{Ambjorn:2011cg,Ambjorn:2012jv}. The case of Euclidean dynamical triangulations is under renewed investigation \cite{Laiho:2016nlp}.
The FRG is a suitable tool to complement such simulations and discover candidates for a universal continuum limit beyond branched polymers.

We consider a complex rank-4 tensor model, i.e., we work with 
a
random tensor $T_{abcd}$ and its complex conjugate $\bar{T}_{abcd}$ of size $N^\prime$. 
We focus on a
model 
respecting the following symmetry
\begin{eqnarray}
T_{a_1 a_2 a_3 a_4}\,&\rightarrow&\, T^\prime _{a_1 a_2 a_3 a_4} = U^{(1)}_{a_1 b_1}U^{(2)}_{a_2 b_2}U^{(3)}_{a_3 b_3}U^{(4)}_{a_4 b_4} T_{b_1 b_2 b_3 b_4}\,,\nonumber\\
\bar{T}_{a_1 a_2 a_3 a_4}\,&\rightarrow&\, \bar{T}^\prime _{a_1 a_2 a_3 a_4} = \bar{U}^{(1)}_{a_1 b_1}\bar{U}^{(2)}_{a_2 b_2}\bar{U}^{(3)}_{a_3 b_3}\bar{U}^{(4)}_{a_4 b_4} \bar{T}_{b_1 b_2 b_3 b_4}\,,
\label{4d1}
\end{eqnarray}
where repeated indices are summed over. The matrices $U^{(i)}_{ab}$ are unitary and therefore the model has a $U(N^\prime )^{\otimes 4}$ symmetry. Eq.~\eqref{4d1} shows that each index of the tensor transforms independently. 
Hence,
$U(N^\prime )^{\otimes 4}$ invariance requires that the only  
allowed
index contraction is a first index of $T$ with a first index of $\bar{T}$, a second index of $T$ with a second index of $\bar{T}$ and so on. Consequently, an interaction term which contains $2p$ tensors in total necessarily has $p$ tensors $T$ and $p$ tensors $\bar{T}$. Invariance under Eq.~\eqref{4d1} also ensures that the indices of the tensors do not have any permutation symmetry. 

\begin{figure}[t]
\begin{center}
\includegraphics[scale=.4]{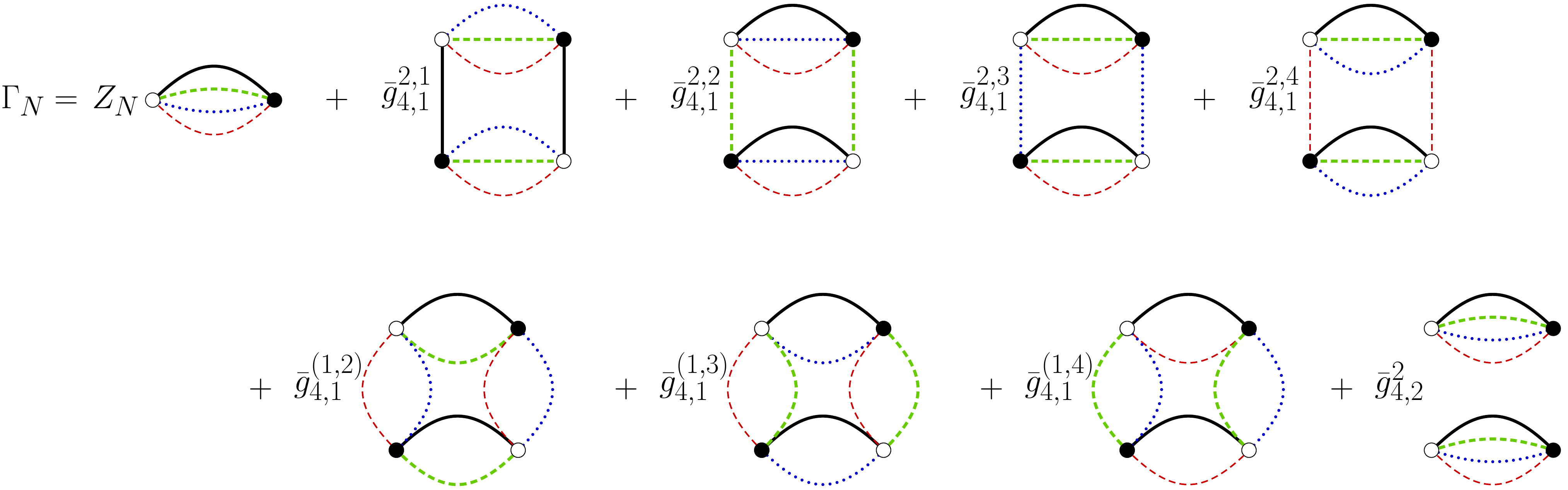}
\caption{Diagrammatic representation of the invariants present in our truncation}
\label{trunc}
\end{center}
\end{figure}

A
continuum limit in tensor models might fall into the universality class corresponding to the Reuter fixed point  \cite{Reuter:1996cp} (see \cite{Eichhorn:2017egq,Eichhorn:2018yfc} for recent reviews).  Accordingly we
bootstrap our truncation assuming a near-canonical scaling spectrum, and choose
\begin{equation}
\Gamma_N = \Gamma_{N,2} + \Gamma_{N,4}\,,
\label{4d3}
\end{equation}
with
\begin{equation}
\Gamma_{N,2} = Z_N \bar{T}_{a_1 a_2 a_3 a_4} T_{a_1 a_2 a_3 a_4}\,,
\label{4d4}
\end{equation}
and
\begin{eqnarray}
\Gamma_{N,4} &=&  \bar{g}^{2,1}_{4,1}\, \bar{T}_{a_1 a_2 a_3 a_4} T_{b_1 a_2 a_3 a_4}\bar{T}_{b_1 b_2 b_3 b_4}T_{a_1 b_2 b_3 b_4}+ \bar{g}^{2,2}_{4,1}\, \bar{T}_{a_1 a_2 a_3 a_4} T_{a_1 b_2 a_3 a_4}\bar{T}_{b_1 b_2 b_3 b_4}T_{b_1 a_2 b_3 b_4}\nonumber\\
&+&  \bar{g}^{2,3}_{4,1}\, \bar{T}_{a_1 a_2 a_3 a_4} T_{a_1 a_2 b_3 a_4}\bar{T}_{b_1 b_2 b_3 b_4}T_{b_1 b_2 a_3 b_4}+ \bar{g}^{2,4}_{4,1} \,\bar{T}_{a_1 a_2 a_3 a_4} T_{a_1 a_2 a_3 b_4}\bar{T}_{b_1 b_2 b_3 b_4}T_{b_1 b_2 b_3 a_4}\nonumber\\
&+& \bar{g}^{(1,2)}_{4,1}\, \bar{T}_{a_1 a_2 a_3 a_4} T_{b_1 b_2 a_3 a_4}\bar{T}_{b_1 b_2 b_3 b_4} T_{a_1 a_2 b_3 b_4} + \bar{g}^{(1,3)}_{4,1}\, \bar{T}_{a_1 a_2 a_3 a_4} T_{b_1 a_2 b_3 a_4}\bar{T}_{b_1 b_2 b_3 b_4} T_{a_1 b_2 a_3 b_4}\nonumber\\ 
&+&\bar{g}^{(1,4)}_{4,1}\, \bar{T}_{a_1 a_2 a_3 a_4} T_{b_1 a_2 a_3 b_4}\bar{T}_{b_1 b_2 b_3 b_4} T_{a_1 b_2 b_3 a_4}+\bar{g}^{2}_{4,2}\, \bar{T}_{a_1 a_2 a_3 a_4} T_{a_1 a_2 a_3 a_4} \bar{T}_{b_1 b_2 b_3 b_4}T_{b_1 b_2 b_3 b_4}\,.\nonumber\\
\label{4d5}
\end{eqnarray}
 In Eq.~\eqref{4d4}, $Z_N$ denotes the wave-function renormalization. 
 The interactions in Eq.~\eqref{4d5} can be represented by 4-colored graphs:  
 For each tensor $T$ ($\bar{T}$) we associate a white (black) vertex  
 and
 a colored edge for an index. Different colors are used to indicate 
 different index positions on the tensors. An index contraction is represented by linking 
 a black and a white vertex
 by the corresponding edge. 
The corresponding
 diagrammatic representation of Eq.~\eqref{4d3} is shown in Fig.~\ref{trunc}. 
 The notation for the couplings of different interactions encode the corresponding diagrammatics, i.e., the combinatorial structures of the interaction: The first subindex denotes the number of tensors $T$ and $\bar{T}$, while the second one counts the number of connected components.
The
superindices differ for different combinatorial structures. For cyclic melons, which consist of contractions of neighboring tensors by either three or one line in alternating fashion, the superindices are not bracketed. The two superindices
stand for the number of ``submelons" and the preferred color $i$ which is the color of the single line connecting neighboring tensors. Since this is a rank-4 model, there are four different melonic invariants, each one selecting one distinct
preferred color. A symmetry-reduced theory space, the isocolored theory space is defined by a single coupling being assigned to all cyclic melons since those interactions have the same combinatorial structure and just differ by the preferred color. \\
A distinct combinatorial structure is indicated by bracketed superindices:
The couplings $\bar{g}^{(1,i)}_{4,1}$ are associated to the \textit{necklaces} diagrams. These interactions are such that a given white vertex is connected to a black vertex by exactly two edges. There are three such interactions due to three possible pairings of four indices into two groups of two. Each white vertex is connected to one of its neighbors by the colors $(1,i)$ in the superindex, and by the remaining two colors to its other neighbor. \\
Finally,  
the ``double-trace" interaction  
is parameterized by the
coupling 
$\bar{g}^{2}_{4,2}$: 
in the subindices, 
the number of connected components 
is
two. 
The superindex represents the number of melons. 
At higher orders in the truncation, where the first subindex is larger than four, additional superindices must be introduced to distinguish all different combinatorial structures at fixed order in tensors and connected components.

We aim at deriving the beta functions for the dimensionless couplings $g_I$, where 
\be
\bar{g}_I\equiv Z^2_N N^{[\bar{g}_I]} g_I,
\ee
 with $[\bar{g}_I]$ being the canonical dimension of the coupling $\bar{g}_I$. 
 We will now determine
 the canonical dimensions, cf.~Sec.~\ref{sec:Scalingdim}.
 For the application of the flow equation one has to choose a regulator function which acts as an ``infrared" suppression term, cutting off modes with indices satisfying $a^p+b^p+c^p+d^p < N^r$,  
where
 $r,p >0$. Our regulator choice 
 generalizes that in \cite{Eichhorn:2018ylk} 
\begin{equation}
R^{(r,p)}_N (\left\{a_i\right\},\left\{b_i\right\}) = Z_N \,\delta_{a_1 b_1}\delta_{a_2 b_2}\delta_{a_3 b_3}\delta_{a_4 b_4}\left(\frac{N^r}{\sum^4_{i=1} a^p_i}-1\right)\theta\left(\frac{N^r}{\sum^4_{j=1} a^p_j}-1\right)\,.
\label{4d6}
\end{equation}
Its scale-derivative $\partial_t \equiv N\partial_N$ is
\bea
\partial_t R^{(r,p)}_N (\left\{a_i\right\},\left\{b_i\right\}) &=&  \delta_{a_1 b_1}\delta_{a_2 b_2}\delta_{a_3 b_3}\delta_{a_4 b_4}\left[N^r\frac{\partial_t Z_N + Z_N}{\sum^4_{i=1} a^p_i}-\partial_t Z_N\right]\theta\left(\frac{N^r}{\sum^4_{j=1} a^p_j}-1\right) \label{eq:Rdot}\\
&{}& + Z_N \,\delta_{a_1 b_1}\delta_{a_2 b_2}\delta_{a_3 b_3}\delta_{a_4 b_4}\left(\frac{N^r}{\sum^4_{i=1} a^p_i}-1\right)\cdot \delta \left( \frac{N^r}{\sum^4_{i=1} a^p_i}-1\right) \cdot r\,\frac{N^r}{\sum^4_{k=1} a^p_k}\,.\nonumber
\eea
The term in the second line of Eq.~\eqref{eq:Rdot} does not yield a contribution to the flow of couplings of index-independent interactions, since the delta-distribution appears multiplied by its argument.
With these definitions, the right-hand-side of the flow equation can be evaluated. To extract beta functions from it, suitable projections onto the monomials spanning the theory space have to be used. Specifically, the distinct combinatorial structures at a given order in tensors can easily be distinguished, as the flow equation generates combinatorially different contractions on the right-hand-side. To deal with the additional index-dependence of interactions that occurs due to the symmetry breaking induced by the regulator, we apply the prescription from \cite{Eichhorn:2018ylk}. Specifically, the regulator can either sit on an index forming a closed loop, or an index occurring on a tensor and antitensor. To project onto symmetry-invariant monomials only, we set indices in the regulator to zero, if they also occur on a tensor. This splits the index-trace into two parts: The contraction of tensors and their complex conjugates
decouples from the regulator trace and is directly recognizable as one of the different combinatorial structures in Eq.~\eqref{4d4} or \eqref{4d5}. The regulator trace consists of a trace over indices running through the regulator and its derivative, which can be rewritten as an integral in the large-$N$ limit.\\ 
The resulting beta functions for the dimensionless couplings as well as the anomalous dimension $\eta \equiv -\partial_t Z_N /Z_N$ are, respectively, 
\bea
\eta &=& 2\,\EuScript{I}^{3}_2 (p) N^{[\bar{g}^{2,i}_{4,1}]+\frac{3r}{p}} \sum^{4}_{i=1}{g}^{2,i}_{4,1}+4\,\EuScript{I}^{2}_2 (p) N^{[\bar{g}^{(1,i)}_{4,1}]+\frac{2r}{p}}\sum^{4}_{i=2}{g}^{(1,i)}_{4,1}+2\,\EuScript{I}^{4}_2 (p) N^{[\bar{g}^2_{4,2}]+\frac{4r}{p}} {g}^2_{4,2}\,,
\label{anomdim1}\\
\beta_{g^{2,i}_{4,1}} &=& \left(2\eta - [\bar{g}^{2,i}_{4,1}]\right)g^{2,i}_{4,1}+4\,\EuScript{I}^{3}_3 (p) N^{[\bar{g}^{2,i}_{4,1}]+\frac{3r}{p}}(g^{2,i}_{4,1})^2+8\,\EuScript{I}^{1}_3 (p) N^{2[\bar{g}^{(1,i)}_{4,1}]-[\bar{g}^{2,i}_{4,1}]+\frac{r}{p}}\sum^{4}_{j=2}\sum_{k>j}g^{(1,j)}_{4,1}g^{(1,k)}_{4,1}\nonumber\\
&+& 8\,\EuScript{I}^{2}_3 (p) N^{[\bar{g}^{(1,i)}_{4,1}]+\frac{2r}{p}}g^{2,i}_{4,1}\sum^{4}_{j=2}g^{(1,j)}_{4,1}\,,
\label{cyclimelon1}\\
\beta_{g^{(1,i)}_{4,1}}& =& \left(2\eta -[\bar{g}^{(1,i)}_{4,1}]\right) g^{(1,i)}_{4,1}+8\,\EuScript{I}^{2}_3 (p)N^{[\bar{g}^{(1,i)}_{4,1}]+\frac{2r}{p}}(g^{(1,i)}_{4,1})^2+8\,\EuScript{I}^{1}_3 (p)N^{[\bar{g}^{2,i}_{4,1}]+\frac{r}{p}}g^{(1,i)}_{4,1}\sum^4_{j=1}g^{2,j}_{4,1}\,,\\
\label{necklaces1}\\
\beta_{g^{2}_{4,2}}&=&\left(2\eta-[\bar{g}^{2}_{4,2}]\right)g^{2}_{4,2}+8\,\EuScript{I}^{2}_3 (p)N^{2[\bar{g}^{2,i}_{4,1}]-[\bar{g}^{2}_{4,2}]+\frac{2r}{p}}\sum^4_{i=1}\sum_{j>i}g^{2,i}_{4,1}g^{2,j}_{4,1}\nonumber\\
&+&8\,\EuScript{I}^{1}_{3}(p) N^{[\bar{g}^{2,i}_{4,1}]+[\bar{g}^{(1,i)}_{4,1}]-[\bar{g}^{2}_{4,2}]+\frac{r}{p}}\sum^4_{i=1}\sum^4_{j=2}g^{2,i}_{4,1}g^{(1,j)}_{4,1}+8\,\EuScript{I}^{3}_3 (p)N^{[\bar{g}^{2,i}_{4,1}]+\frac{3r}{p}}g^{2}_{4,2}\sum^4_{i=1}g^{2,i}_{4,1}\nonumber\\
&+&16\,\EuScript{I}^{2}_3 (p)N^{[\bar{g}^{(1,i)}_{4,1}]+\frac{2r}{p}}g^{2}_{4,2}\sum^4_{i=2}g^{(1,i)}_{4,1}+4\,\EuScript{I}^{4}_3 (p)N^{[\bar{g}^{2}_{4,2}]+\frac{4r}{p}}(g^{2}_{4,2})^2\,,
\label{doubletrace1}
\end{eqnarray}
where $\EuScript{I}^{i}_j (p)$ are threshold integrals 
provided in
App.~\ref{thresholdintegrals} for $p=1,2$. \\
The canonical dimensions for the couplings are not fixed in Eq.~\eqref{anomdim1} to \eqref{doubletrace1}. 
They are fixed by demanding that 
Eqs.~\eqref{anomdim1}-\eqref{doubletrace1} admit a $1/N$ expansion 
starting with
a non-trivial contribution at order $(1/N)^0$. In
the expression for the anomalous dimension, Eq.~\eqref{anomdim1}, the large-$N$ limit can be taken if the canonical dimensions satisfy the following bounds
\begin{equation}
\frac{3r}{p}+[\bar{g}^{2,i}_{4,1}]\leq 0\,,\qquad \frac{2r}{p}+[\bar{g}^{(1,i)}_{4,1}]\leq 0\,,\qquad \frac{4r}{p}+[\bar{g}^{2}_{4,2}]\leq 0\,.
\label{candim1}
\end{equation}
Eq.~\eqref{cyclimelon1} imposes a new constraint for the canonical dimensions, namely
\begin{equation}
2[\bar{g}^{(1,i)}_{4,1}]-[\bar{g}^{2,i}_{4,1}]+\frac{r}{p}\leq 0\,,
\label{candim2}
\end{equation}
while the beta function for the necklaces \eqref{necklaces1} does not introduce any new conditions. Finally, the beta function for the double-trace couplings constrains the canonical dimensions by
\begin{equation}
\frac{2r}{p}+2[\bar{g}^{2,i}_{4,1}]-[\bar{g}^2_{4,2}]\leq 0\qquad\mathrm{and}\qquad[\bar{g}^{2,i}_{4,1}]+[\bar{g}^{(1,i)}_{4,1}]-[\bar{g}^2_{4,2}]+\frac{r}{p}\leq 0\,.
\label{candim3}
\end{equation}
From Eqs.~\eqref{candim1}, \eqref{candim2} and \eqref{candim3}, we obtain
upper bounds for the couplings canonical dimensions (or relations between them). Couplings decouple from the set of beta functions if their canonical dimension is chosen below the corresponding upper bounds. In this sense choosing the upper bounds as the canonical dimensions leads to the most non-trivial set of beta functions at large $N$. We tentatively consider a decoupling of interactions through such choices artificial,  
hence, we
choose the upper bounds as the canonical dimension for the couplings. 
We start by demanding that 
\begin{equation}
[\bar{g}^{2,i}_{4,1}] = -\frac{3r}{p}\,.
\label{candimcycmel}
\end{equation}   
This is exactly what one would expect based on $[\bar{g}^{2,i}_{4,1}]_{{\rm rank}-3} = -2r$ (for $p=1$) in the rank-3 case: The contraction of one additional index in the rank-4 case requires an additional suppression by $1/N$ (for $r/p=1$).
The first inequality of \eqref{candim3} implies $[\bar{g}^{2}_{4,2}]\geq -4r/p$. Yet,
the third inequality of \eqref{candim1} enforces
$[\bar{g}^{2}_{4,2}]\leq -4r/p$. Therefore, given the choice in Eq.~\eqref{candimcycmel}, the canonical dimension for the double-trace coupling is completely fixed, i.e.,
\begin{equation}
[\bar{g}^{2}_{4,2}] = -\frac{4r}{p}\,.
\label{candimdoubletrace}
\end{equation}
This is in accordance with the expectation from the rank-3 case, as well as the reasoning that the additional ``trace" of this interaction in comparison to the quartic cyclic melonic interaction should lead to an additional suppression by $1/N$ (for $r=p$).
Finally, using \eqref{candimcycmel} and \eqref{candimdoubletrace} in \eqref{candim1}-\eqref{candim3} results in $[\bar{g}^{(1,i)}_{4,1}]\leq -2r/p$, i.e., the canonical dimension for the necklaces is not fixed uniquely.
We choose
\begin{equation}
[\bar{g}^{(1,i)}_{4,1}] = -\frac{2r}{p}\,.
\label{candimnecklaces}
\end{equation}
The scaling dimensions are functions of the ratio $r/p$. Hence, if one chooses the ``standard" scaling, i.e., setting the power of the infrared cutoff $N$ equal to that of 
the ``momentum" scale, $r=p$, the dimensions are always $-3$, $-4$ and $-2$ for the cyclic melons, multitrace and necklaces interactions, respectively. For those fixed points that do not feature dimensional reduction, the choice $r=p$ is preferred based on a geometrical argument, see Sec.~\ref{sec:Scalingdim}.
 Nevertheless, the threshold integrals $\EuScript{I}^{i}_j$ depend on those parameters in a non-trivial way which implies that different choices of $(r,p)$ lead to different numerical coefficients in the beta functions. 
Thus,
 choosing different values $r$ and $p$ while keeping  
 all canonical dimensions fixed tests the scheme/regulator dependence of our calculation. \\
In the large-$N$ limit and using Eq.~\eqref{candimcycmel}-\eqref{candimnecklaces} with $r=p=1$, the system of beta functions reduces
to 
\begin{equation}
\eta = \frac{1}{20}(5-\eta)\sum^4_{i=1} g^{2,i}_{4,1}+\frac{1}{3}(4-\eta)\sum^4_{i=2} g^{(1,i)}_{4,1}+\frac{1}{90}(6-\eta)g^2_{4,2}\,,
\label{eta1}
\end{equation}
which can be solved for $\eta$ leading to 
\begin{equation}
\eta = \frac{3\left(15\sum^4_{i=1}g^{2,i}_{4,1}+80\sum^{4}_{i=2}g^{(1,i)}_{4,1}+4g^2_{4,2}\right)}{180+9\sum^{4}_{j=1}g^{2,i}_{4,1}+60\sum^4_{j=2}g^{(1,j)}_{4,1}+2g^2_{4,2}}\,,
\label{eta2}
\end{equation}
and
\begin{eqnarray}
\beta_{g^{2,i}_{4,1}}& =& \left(2\eta +3\right)g^{2,i}_{4,1}+\frac{1}{15}(6-\eta)(g^{2,i}_{4,1})^2+\frac{2}{3}(4-\eta)\sum^{4}_{j=2}\sum_{k>j}g^{(1,j)}_{4,1}g^{(1,k)}_{4,1}
+\frac{2}{5}(5-\eta)g^{2,i}_{4,1}\sum^{4}_{j=2}g^{(1,j)}_{4,1}\,,
\label{g2i41beta}\\
\beta_{g^{(1,i)}_{4,1}} &=& \left(2\eta +2\right) g^{(1,i)}_{4,1}+\frac{2}{5}(5-\eta)(g^{(1,i)}_{4,1})^2\,,
\label{g1i41beta}\\
\beta_{g^{2}_{4,2}}&=&\left(2\eta+4\right)g^{2}_{4,2}+\frac{2}{5}(5-\eta)\sum^4_{i=1}\sum_{j>i}g^{2,i}_{4,1}g^{2,j}_{4,1}+\frac{2}{3}(4-\eta)\sum^4_{i=1}\sum^4_{j=2}g^{2,i}_{4,1}g^{(1,j)}_{4,1}\nonumber\\
&+&\frac{2}{15}(6-\eta)g^{2}_{4,2}\sum^4_{i=1}g^{2,i}_{4,1}+\frac{4}{5}(5-\eta)g^{2}_{4,2}\sum^4_{i=2}g^{(1,i)}_{4,1}+\frac{1}{63}(7-\eta)(g^{2}_{4,2})^2\,.
\label{g242beta}
\end{eqnarray}
We highlight several key features of the above system:
Firstly, unlike in the quartic truncation for the rank-3 complex tensor model \cite{Eichhorn:2017xhy}, there is a class of interactions, the necklaces, which are not melonic.  
On the other hand,  
in
the real rank-3 tensor model, see \cite{Eichhorn:2018ylk}, a non-melonic interaction is present already at the quartic order. It 
does not contribute to the anomalous dimension at large $N$. As a difference to these two examples, 
Eq.~\eqref{eta1} and \eqref{eta2} show that all couplings contribute to the anomalous dimension in the complex rank-4 model, including the non-melonic (necklaces) couplings. This is the first evident structural difference between the present beta functions and the rank-3 ones \cite{Eichhorn:2017xhy,Eichhorn:2018ylk}. \\
Secondly, after choosing the canonical dimension for the melonic coupling $\bar{g}^{2,i}_{4,1}$, the canonical dimension for the double-trace coupling is fixed uniquely.
Its value differs from the canonical dimension of the melonic coupling. Consequently, 
interactions which contain the same number of fields (tensors) as well as sums over indices (which are
the analogue of an integral over momenta in ordinary quantum field theories on a background) scale with different powers at large $N$. This is an intrinsic property of the combinatorially non-trivial structure of the interactions in tensor models, see also  \cite{Eichhorn:2017xhy,BenGeloun:2018ekd,Eichhorn:2018ylk}. 
We caution that
if the double-trace interaction was not introduced in the present truncation, one could choose the canonical dimension for the necklaces to be the same as the canonical dimension of the melonic coupling. This would lead to the misleading conclusion that is 
possible to  
choose
the same canonical dimension for all interactions with a given number of tensors.

We look for fixed points of
the system of beta functions 
Eq.~\eqref{eta2}-\eqref{g242beta}. The strategy is the same as the one employed in \cite{Eichhorn:2017xhy,Eichhorn:2018ylk}: Firstly, zeros of the beta functions are obtained in a perturbative approximation, i.e., the anomalous dimension is taken as a polynomial function of the couplings,  
\begin{equation}
\eta_p = \frac{1}{4}\sum^4_{i=1} g^{2,i}_{4,1}+\frac{4}{3}\sum^4_{i=2} g^{(1,i)}_{4,1}+\frac{1}{15}g^2_{4,2}\,.
\label{etapert}
\end{equation}
With Eq.~\eqref{etapert}, the beta functions are polynomials in the couplings. Hence
finding their zeros is easily achieved with computer software. Once the zeros are obtained, several criteria are applied to filter out candidates for physical fixed points. These include the regulator bound\footnote{ As discussed in \cite{Meibohm:2015twa} and adapted to tensor models in \cite{Eichhorn:2018ylk}, the condition that the regulator diverges at $N \to N' \to \infty$ imposes a bound on the anomalous dimension. For our regulator choice Eq.~\eqref{4d6} this is $\eta<r$.}
$\eta_p < 1$. Further, the critical exponents should stay bounded such that the bootstrap strategy for the choice of truncation is justified. Finally, we demand stability under extensions of the truncation. Given the limited nature of our investigation for the purposes of this review, the only extension is that from the perturbative form of the anomalous dimension in Eq.~\eqref{etapert} to the full expression in Eq.~\eqref{eta2}.

The resulting candidates for physical universality classes
can be separated into two main classes: those with enhancement of the
$U(N^\prime)^{\otimes 4}$ symmetry to $U(N'^2)\otimes U(N^\prime)^{\otimes 2}$, $U(N'^3)\otimes U(N^\prime)$ or $U(N'^4)$ 
display dimensional reduction, i.e., the associated continuum limit would not correspond to four-dimensional geometries. In contrast, 
those with $U(N^\prime)^{\otimes 4}$ symmetry might be 
possible candidates for a suitable continuum limit which could correspond to 4d quantum gravity. The following results  
are quoted for the case
$r=p=1$ unless stated otherwise.

\subsection{Symmetry-enhanced fixed points: Dimensional reduction in tensor models}
The $U(N^\prime)^{\otimes 4}$ symmetric theory space contains 
 symmetry-enhanced subspaces, such as, e.g., 
 $U(N'^2)\otimes U(N')^{\otimes 2}$. 
To achieve the corresponding enhancement of symmetry, interactions which violate it have to be switched off. This happens at several fixed points in our truncation \footnote{Note that while symmetry-enhanced subspaces are invariant under the flow, this does not automatically imply that they must feature interacting fixed points; it can also be the case that the symmetry-enhanced hypersurface is a fixed surface of the flow.}. 
The enhanced symmetry is broken if there is at least one nonvanishing interaction for each of the four colors that treats this color differently form the remaining colors.
Therefore, although it appears slightly paradoxically at first glance,
fixed points which are \emph{not}
 invariant under color permutations typically feature 
a larger symmetry
 than $U(N^\prime)^{\otimes 4}$. The breaking of the color permutation symmetry at the fixed point \footnote{The color permutation symmetry is still intact in the full theory space, as each such fixed point automatically comes with partners related
 by a color permutation.} allows for some interactions to vanish such that a pair, or even triple, of indices can be summarized into one ``superindex". This superindex 
 features an $U(N'^2)$ (or even $U(N'^3)$) symmetry.
Consequently, the interactions which are turned on at the fixed point can be described by lower rank tensors. 
This is a form of dimensional reduction, i.e., the lower rank tensors ``tessellate" lower-dimensional discrete geometries. For instance, at a fixed point at which two pairs of indices are summarized into two superindices, the rank-4-model reduces to a matrix model, which encodes random geometries in two dimensions. 

The enhancement in symmetry and dimensional reduction entails that universality classes of lower-rank-models can be reproduced. Two comments are in order here:\\
Firstly, the recovery of ``lower-dimensional" universality classes requires to exploit the freedom in the choice of regulator in Eq.~\eqref{4d6} such that the canonical dimensions of the interactions agree with those of the lower-rank-model. For instance, the quartic cyclic melonic couplings have canonical dimension $-3 r/p$ in the rank-4- case and $-2$ in the rank -3 case for $r=p=1$. Choosing $r/p=2/3$ for the rank-4-case leads to an agreement in the canonical dimension of the cyclic melons. Analogous choices for different fixed points will be spelled out below. We emphasize that the choice of canonical dimension of quartic interactions is grounded in geometric arguments. Therefore, for each dimensionality $d$, there is a unique choice of $r/p$ for each rank $n$, such that the canonical scaling exponents agree with those required for an identification of the dual of the tensor model with random geometries in $d$ dimensions. This choice appears to be $r/p=1$ for $n=d$, but differs if $n \neq d$.\\ 
Secondly, the symmetry-enhanced fixed points are embedded in a larger theory space with symmetry-breaking directions. Therefore, additional relevant directions might exist which entail additional tuning required to reach the fixed point. Similar enlargements of universality classes are well known in statistical physics. For instance, the scaling exponents for the $O(N+M)$ Wilson-Fisher fixed point can be recovered within a $O(N)\oplus O(M)$ symmetric theory space. Yet, an additional relevant direction is associated to the additional tuning required to reach this critical point, see, e.g., \cite{Eichhorn:2013zza}. We will check on a case-by-case basis whether dimensional reduction requires additional tuning, or whether it is a preferred IR-endpoint of tensorial RG flows.

The 
set of beta functions given by Eq.~\eqref{eta2}-\eqref{g242beta} admits the following symmetry-enhanced fixed points:
\begin{itemize}
\item {\it Cyclic-Melonic Single-trace Fixed Point:} Only one representative of the cyclic melonic interactions $g^{2,i}_{4,1}$ is non-vanishing at this fixed point. For a given cyclic melon, e.g., $g^{2,1}_{4,1}$, the interaction can be expressed as
\begin{equation}
g^{2,1}_{4,1}\bar{T}_{a_1 a_2 a_3 a_4}T_{b_1 a_{2}a_{3} a_4}\bar{T}_{b_{1}b_2 b_3 b_4}T_{a_1b_2 b_3 b_4 }\,\longrightarrow\, g^{2,1}_{4,1}\bar{T}_{a_1 I}T_{b_1 I}\bar{T}_{b_{1}J}T_{a_1J }\,,
\label{cmstfp1}
\end{equation}
where the super-index $I$ condenses three of the initial indices and thereby enhances the symmetry of the model to $U(N^\prime)\otimes U({N^\prime}^ 3)$. Consequently, the fixed-point dynamics is described by a single-matrix model. The corresponding
continuum limit is not
associated with 4d quantum gravity but rather expected to yield the well-known pure-gravity scaling exponent in
2d. 
For $r=p=1$ this fixed point features two relevant directions in 
our simple
truncation, $\theta_1 = 3.47$ and $\theta_2 = 0.31$. 
Due to the systematic error associated with the truncation  
the present results are insufficient to
establish whether the second relevant direction turns into an irrelevant one. We provide a rough estimate for a lower bound on the systematic error by exploiting the freedom in the shape function:
Considering, for instance, a ``spherical" cutoff function, i.e., $r=p=2$, this fixed point also displays two relevant directions with critical exponents $\theta_1 = 3.71$ and $\theta_2 = 0.22$. 

Although associated with a matrix model, the critical exponents reported are far from the exact result obtained for 
the pure-gravity scaling exponent in 2d,
$(\theta = 0.8)$. This is similar to the result obtained in the rank-3 real model in \cite{Eichhorn:2018ylk} and a consequence of 
the canonical dimensional of the cyclic melonic coupling for $r/p=1$. Instead setting $r=1/3$ for $p=1$ implies $[\bar{g}^{2,i}_{4,1}]=-1$ in agreement with the canonical dimension of the quartic interaction in matrix models. In this case, the fixed point has two relevant directions with critical exponents $\theta_1 = 1.05$ and $\theta_2 = 0.11$. For the prescription for critical exponents reported in Sec.~\ref{sec:matrixmodel}, we obtain  $\tilde{\theta}_1 = 0.44$ and $\tilde{\theta}_2 = 0.11$.
More sophisticated truncations are necessary to establish whether the second critical exponents is indeed positive.

\begin{figure}[t]
  \includegraphics[width=\linewidth]{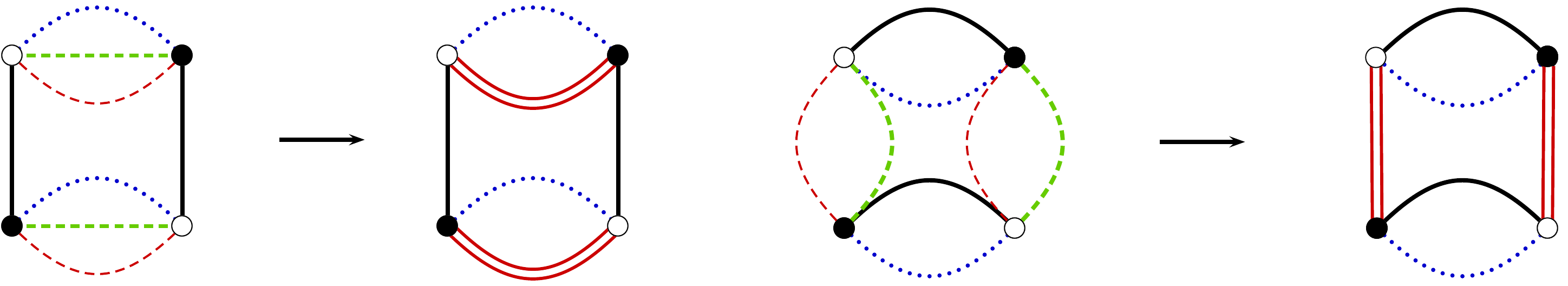}
  \caption{Illustration of symmetry enhancement: At a fixed point which features the cyclic melon with coupling $g_{4,1}^{2,1}$ and the necklace with coupling $g_{4,1}^{(1,2)}$, two indices, represented by green and red dashed lines are collected in one red double-line which represents a super-index. At a fixed point which only features the necklace $g_{4,1}^{(1,2)}$, the index pair $(1,2)$ as well as the pair $(3,4)$ can be summarized to two superindices, entailing a reduction to a matrix model.}
  \label{rank-3dimred}
\end{figure}

\item {\it Multitrace-Bubble Fixed Point:} All interactions but the double-trace $g^{2}_{4,2}$ one vanish
at this fixed point. The remaining interaction can be expressed as 
\begin{equation}
g^{2}_{4,2}\bar{T}_{a_1 a_2 a_3 a_4}T_{a_1 a_{2}a_{3} a_4}\bar{T}_{b_{1}b_2 b_3 b_4}T_{b_1b_2 b_3 b_4 }\,\longrightarrow\, g^{2}_{4,2}\bar{T}_{I}T_{I}\bar{T}_{J}T_{J}\,,
\label{mbfp1}
\end{equation}
where all indices  
are
collected in one single super-index $I$. Such a term is characterized by an
enhanced symmetry $U({N^\prime}^ 4)$ and it describes a vector model. This fixed point displays four relevant directions: $\theta_1=4.69$ and $\theta_{2,3,4}=0.20$. The three-fold degeneracy is a consequence of an exchange-symmetry between the three directions that break the enhanced symmetry. 
The small absolute value of $\theta_{2,3,4}$ does not permit to determine whether there are four relevant directions in total. In fact,
the same fixed-point structure is seen in the rank-3 model \cite{Eichhorn:2018ylk}: 
there,
it features two relevant directions in the quartic truncation while in the hexic truncation, only one relevant direction remains, see \cite{Eichhorn:2018ylk}. 

\item {\it  
Single-necklace
Fixed Point:} Only one necklace interaction is non-vanishing at this fixed point. All other interactions in our truncation vanish. Beyond the truncation, only interactions respecting the corresponding enhanced symmetry are present.
Due to the color-permutation symmetry in the theory space, there are three such fixed points, each characterized by a different non-vanishing necklace.
If one takes, e.g., $g^{(1,2)}_{4,1}$ to be the non-vanishing necklace at the fixed point, the interaction term can be expressed as
\begin{equation}
g^{(1,2)}_{4,1}\bar{T}_{a_1 a_2 a_3 a_4}T_{a_1 a_{2} b_{3} b_4}\bar{T}_{b_{1}b_2 b_3 b_4}T_{b_1b_2 a_3 a_4 }\,\longrightarrow\, g^{(1,2)}_{4,1}\bar{T}_{IJ}T_{IK}\bar{T}_{LK}T_{LJ}\,,
\label{pnfp1}
\end{equation}
where the two 
index pairs are
collected in one super-index $I$. Therefore, the interaction features an enhanced $U({N^\prime}^2)\otimes U({N^\prime}^2)$ symmetry. Accordingly, it
describes a matrix model with matrices $(\bar{T},{T})$. This
symmetry enhancement is associated with an effective dimensional reduction: the continuum limit associated with this fixed point does
not correspond to 4d quantum gravity, but 2d random geometries. This fixed point  has one relevant direction with critical exponent $\theta = 2.27$. The naive value of the universal scaling exponent 
deviates strongly from the exact result $\theta = 0.8$. We
attribute the difference to the fact that the canonical dimension for the necklaces couplings is $-2$ and not $-1$ as would be in the assignment of the canonical dimensions in matrix models, see \cite{Eichhorn:2013isa}. By choosing $r=1/2$ for $p=1$, the fixed point exhibits one relevant direction with scaling exponent $\theta = 1.07$ which gets closer to the exact result. The second prescription for the universal scaling exponents yields  $\tilde{\theta}=0.42$.\\
As a simple check of the robustness of these results, we explore the choice $r=p=2$. We obtain one
relevant direction with critical exponent $\theta=2.37$. Assigning dimension $-1$ for $p=2$ requires $r=1$. For this choice, the fixed point has one relevant direction with critical exponent
$\theta=1.09$. The results are qualitative and even numerically compatible with those discussed for $p=1$, giving a first hint towards stability under different choices of scheme.
\end{itemize}

The above fixed points are characterized by
a single
interaction type. Symmetry-enhanced fixed points with more than one non-vanishing interaction are also possible. These include, e.g.,
\begin{itemize}
\item {\it One Cyclic-Melonic Multitrace Fixed Point:} At this fixed point, just one cyclic melonic interaction of a given preferred color and the double-trace interaction are non-vanishing. If one selects, e.g., the coupling $g^{2,1}_{4,1}$ to be the non-vanishing cyclic melon, the interactions at the fixed-point are given by
\begin{eqnarray}
&& g^{2,1}_{4,1}\bar{T}_{a_1 a_2 a_3 a_4}T_{b_1 a_{2}a_{3} a_4}\bar{T}_{b_{1}b_2 b_3 b_4}T_{a_1b_2 b_3 b_4 }+g^{2}_{4,2}\bar{T}_{a_1 a_2 a_3 a_4}T_{a_1 a_{2}a_{3} a_4}\bar{T}_{b_{1}b_2 b_3 b_4}T_{b_1b_2 b_3 b_4 } \nonumber \\
&{}&\longrightarrow g^{2,1}_{4,1}\bar{T}_{a_1 I}T_{b_1 I}\bar{T}_{b_{1}J}T_{a_1J }+g^{2}_{4,2} \bar{T}_{a_1I}T_{a_1 I}\bar{T}_{b_1 J}T_{b_1 J}\,,
\label{cmmtfp1}
\end{eqnarray}
Three indices  
are
condensed in one super-index $I$, enhancing the symmetry to $U({N^\prime})\otimes U({N^\prime}^ 3)$. Accordingly, the fixed point is associated with a matrix model.  
It features one-relevant direction and the associated critical exponent is $\theta=3.06$. In fact for this case, one cannot reproduce both canonical scaling dimensions for matrix models. To obtain agreement for the single-trace quartic coupling, one should again choose $r/p=1/3$. This yields a canonical dimension of -1 for the single-trace coupling, but -4/3 for the double-trace, whereas the corresponding dimensions in the matrix model are -1 and -2. Therefore it is not clear whether this fixed point admits an interpretation in terms of a matrix model for random geometries.
\end{itemize}

Beyond the fixed-point candidates reported here, further
 zeros of the beta functions 
characterized by
 symmetry enhancement are also obtained. As particular examples, one finds zeros where one cyclic melonic interaction together with one necklace and the multitrace interactions are turned on. Due to the different combinatorial structures,
the corresponding dynamics can be mapped to that of a
 rank-3 model
 as illustrated in Fig.~\ref{rank-3dimred}. 
As the cyclic melons and necklaces feature different canonical dimensions, the corresponding model would presumably be a two-tensor model. A further
 zero of the system of beta functions features a single
 necklace and the double-trace interaction. However, in the present truncation, such zeros of the beta functions violate the regulator bound. Therefore we tentatively discard them and do not consider them as candidates for fixed points, i.e., universal scaling regimes. 
More refined studies are necessary to robustly confirm this characterization.

\subsection{Candidates for four-dimensional emergent space}
In this subsection, we discuss a fixed point which does not feature symmetry enhancement of the form as
discussed previously and therefore cannot be mapped to a lower-rank single tensor model. Thus it might be a potential candidate for the description of 4d quantum gravity. Of course, establishing a universality class for 4d quantum gravity requires much more than just finding a fixed point without dimensional reduction of the form discussed above. After all, the Hausdorff and spectral dimensions as well as further properties of the emergent geometry have not been studied yet.  Nevertheless, the existence of a fixed point that does not admit dimensional reduction to a model of lower rank is most likely a \emph{necessary} requirement for a universality class for 4d quantum gravity. If corroborated by further studies, our discovery might
therefore constitute the very first step on a path towards 4d quantum gravity from tensor models.\\
Here, we
focus on isocolored fixed points, i.e., those that display the same values for all couplings associated with different colors. In other words, we restrict the fixed points to a
symmetry-enhanced subspace which explicitly realizes a discrete color permutation symmetry in all interactions. It is still characterized by the $U(N')^{\otimes 4}$ symmetry and does not feature dimensional reduction to a model of lower rank. 
We conjecture that
in the continuum, color-distinguishing structures should not play any role. This is based on the expectation that color is not associated with a 
physical property of continuum geometries and therefore only isocolored fixed points should matter. We caution that
this might be a naive viewpoint, since the unequal treatment of colors could introduce more sophisticated structures. As stressed before, the identification of universality classes with actual relevance for 4d quantum gravity requires further insights into the emergent geometries. Here, we restrict ourselves to a very first mapping of different fixed-point structures in the theory space. 

\begin{table}[t]
\centering
\begin{tabular}{|c|c|c||c||c|c|c|c|}
\hline
$g^{2,i}_{4,1}$ & $g^{(1,j)}_{4,1}$ & $g^{2}_{4,2}$                            & $\eta$     &$\theta_1$      & $\theta_{2,3}$    & $\theta_{4,5}$                           & $\theta_{6,7,8}$       \\ \hline
$-0.09$         & $-0.07$           & $-5.70$                                  & $-0.91$      &$3.44$          & $0.18$            & \multicolumn{1}{l|}{$-0.40 \pm 0.07\,i$} & $-0.56$        \\ \hline 
\end{tabular}
\caption{Fixed point and critical exponents values for the isocolored fixed point with $r=p=1$}
\label{fpvaluesiso}
\end{table}

In the quartic truncation, one completely isocolored fixed points is found. At this fixed point, all couplings are non-vanishing.  As a consequence, there is no symmetry enhancement of the $U(N')^{\otimes 4}$ symmetry (apart from the discrete color permutation symmetry) which would allow for an immediate identification of dimensional reduction. 
The fixed point as well as the corresponding critical exponents for $r=p=1$ are displayed in Tab.~\ref{fpvaluesiso}:
the isocolored fixed point features three relevant directions. A simple test of the scheme-dependence of this result can be performed
by changing the regulator to $r=p=2$. The isocolored fixed point persists and features positive critical exponents  
$\theta_1=3.41$ and $\theta_{2,3}=0.18$ very close to the values for $r=p=1$.

A subset of these relevant directions could be associated with the tuning towards the isocolored symmetry. To isolate such directions, we re-investigate the fixed point in an isocolored truncation, where 
the different color couplings are identified in Eq.~\eqref{4d5}: $g^{2,1}_{4,1}=g^{2,2}_{4,1}=g^{2,3}_{4,1}=g^{2,4}_{4,1}=g^{2}_{4,1}$ and $g^{(1,2)}_{4,1}=g^{(1,3)}_{4,1}=g^{(1,4)}_{4,1}=g_{4,1}$. The theory space in this truncation is spanned by three couplings. For $r=p=1$, the fixed point displays one relevant direction associated to
$\theta=3.44$. Consequently, the two extra relevant directions that appear in the non-isocolored truncation are associated with an additional
tuning of couplings to achieve the color symmetry at the
fixed point. As discussed above, it remains to be investigated whether color-symmetry breaking can be given any physical meaning in random geometries. Therefore it is currently open whether one should only compare the leading relevant critical exponent $\theta_1=3.44$ to the critical exponents characterizing gravity in the continuum limit, i.e., the critical exponents of the Reuter fixed point, or whether one should also include $\theta_{2,3}$ in the comparison of universality classes.

The isocolored fixed point serves as a prototypical example of a fixed-point structure which does not manifest dimensional reduction at the level of the basic building blocks used to generate random geometries. This is only a necessary condition for a universality class associated to 4d quantum gravity, and
the physical nature of the continuum limit associated to such a fixed point still needs to be investigated. 

Going beyond the isocolored theory space, 
different fixed-point structures than
the completely isocolored one which do not feature symmetry enhancement can be found. In particular, there are fixed points where all couplings are turned on, but for instance, not all couplings of a given combinatorial structure attain the same value. 
These fixed points are not color-permutation invariant. 
A detailed discussion of these new universality classes is beyond the scope of the present review and will be reported in a separate work.

\section{Outlook: Converging to quantum gravity from different directions}\label{sec:conclusions}
We advocate the point of view that an understanding of (key aspects of) quantum gravity can be achieved by making sense of the path integral for quantum gravity. In tensor models, the path integral is interpreted as a sum over random geometries. This sum can be tackled in a dual formulation, where rank $d$ tensors form building blocks of $d$-dimensional space(time). The functional Renormalization Group equation is equivalent to the path integral, as it is simply a way of rewriting an integral into a differential equation that tracks the change of the integral under a change of a parameter in the integrand. This abstract setup translates into the well-known local coarse graining in quantum field theories defined on a background. We highlight that the notion of coarse graining also makes sense in a background-independent setting. In this setting, the number of degrees of freedom provides a background independent notion of scale. In accordance with the intuition behind the a-theorem, the RG flow goes from many to few degrees of freedom. For tensor models, this corresponds to an RG flow in the tensor size $N$. RG fixed points play a crucial role, as they provide universality in the large $N$ limit. Physically, this provides a phase transition in the space of couplings that leads to a continuum phase that is independent of unphysical microscopic details.
 
The path integral for quantum gravity is a point of convergence for a diverse set of
of viewpoints, e.g., \cite{Donoghue:1994dn,Reuter:1996cp,Reuter:2012id,Eichhorn:2018yfc,Daum:2013fu,Feldbrugge:2017kzv,Hamber:2009zz,Perez:2012wv,Ambjorn:1998xu,Ambjorn:2001cv,
 Ambjorn:2012jv,Laiho:2016nlp,Rivasseau:2011hm,Rivasseau:2012yp,Rivasseau:2013uca,Rivasseau:2016zco,Hamber:2009mt,Freidel:2005qe,Baratin:2011hp,Surya:2011yh}. The configuration space that is summed over in these settings typically includes a sum over (discretized) geometries. The inclusion of non-geometric configurations, e.g., \cite{Surya:2011yh} or summation over topologies, is one distinguishing feature of the different approaches that could be of physical relevance. Restricting to the sum over geometries, different approaches to the path integral implement this summation in mathematically distinct ways. These have diverse advantages, such as a direct access to large-scale properties of emergent geometries from lattice simulations, see, e.g., \cite{Ambjorn:2008wc}, a straightforward way of discovering universality classes and characterizing them by their scaling exponents in tensor models \cite{Eichhorn:2017xhy,Eichhorn:2018ylk}, and a direct link to phenomenological questions and the interplay between quantum gravity and matter in the continuum asymptotic safety approach \cite{Eichhorn:2017egq,Eichhorn:2018yfc} to name just a few. We advocate the point of view that such different approaches need not necessarily be considered as competitors in the race towards the goal of discovering quantum gravity. Rather, these different approaches can
 be viewed as different windows that allow us to view and explore distinct aspects of quantum gravity. In the best case, these are complementary, and a comprehensive understanding of quantum spacetime can emerge if key results and strengths from these diverse directions are brought together to form one coherent big picture. As in many other settings, a diversity of viewpoints can accelerate the discovery of a solution to a tough challenge - in this case, quantum gravity. Yet, a diversity of viewpoints brings a new challenge, namely the potential lack of a common language. In quantum gravity, different approaches are often formulated in mathematically very dissimilar ways, making it challenging to extract common physics. Here, we advocate that the functional RG setup could provide one option for a common language shared by
 different approaches. In particular, it allows one to evaluate scaling exponents linked to a universal continuum limit. These universal exponents can be compared, e.g., from continuum asymptotic safety and tensor models \footnote{We stress that if the continuum limit in the gravitational path-integral can be taken, this requires asymptotic safety or asymptotic freedom to hold. In fact, to take the continuum limit in any path integral, while obtaining an interacting, and thus potentially interesting infrared limit, requires one of the two scenarios to hold.
 On the other hand, this does not imply that the universality class need be that of the Reuter fixed point, which is asymptotic safety of a path integral with a particular choice of configuration space. In principle, it might be that the asymptotic-safety scenario is realized in mathematically and physically distinct ways in several different incarnations of the path integral for gravity. It is therefore important to keep in mind the distinction between asymptotic safety as a general scenario and asymptotically safe gravity defined by the Reuter fixed point.}. Once the required precision has been reached in advanced approximation of the full RG flow, such a quantitative comparison will unveil whether these approaches to the gravitational path integral encode the same physics. In the most straightforward setup for tensor models, full agreement with scaling exponents from CDTs or continuum asymptotic safety is probably not expected. This is due to the difference in configuration spaces in the respective approaches to the path integral. For instance, the gluing rules encoding causality in CDTs are expected to lead to a restriction on the allowed (multi)-tensor interactions. Following \cite{Benedetti:2008hc}, the corresponding tensor model, once set up, can be explored by means of the FRG, and a characterization of the universality class is possible.

Understanding the quantum structure of spacetime is a challenging goal. We advocate that the complementarity that different approaches to the path integral for quantum gravity exhibit is a highly promising starting point. 
Tensor models could be helpful in this quest as they could contribute to bridging the gap between discrete numerical and analytical continuum approaches by allowing for a discrete analytical approach.
We propose that background independent functional RG techniques could potentially act as a catalyst for breakthroughs. Specifically, they allow us to discover and characterize universality classes for the continuum limit in tensor models.  This could provide one of the missing links towards a background-independent understanding of quantum gravity.

Even if this hope is not realized, tensor models could constitute a stand-alone approach to the path integral for gravity. To that end, going beyond the simplest form of the large $N$ limit, which leads to a branched-polymer phase, appears to be necessary. We highlight the potential use of the FRG in this context, as it is a highly flexible tool allowing to search for universal scaling regimes in diverse tensor model theory spaces. In particular, setting up truncations adapted to different assumptions regarding the nature of the universality class (e.g., near-canonical vs. fully non-perturbative) could give access to different continuum limits. 

\emph{Acknowledgements}
We acknowledge helpful discussions with V.~Rivasseau,  S.~Carrozza and J.~Ben Geloun.
A.~E.~and A.~D.~P.~are supported by the Deutsche Forschungsgemeinschaft under grant no.~Ei/1037-1. The work of T.~K. was supported in part by the PAPIIT grant no.~IA-103718 at the UNAM.

\appendix

\section{Threshold integrals}\label{thresholdintegrals}

In this appendix we list the expressions for the threshold integrals that appear in the beta functions \eqref{anomdim1}-\eqref{doubletrace1}:
\begin{eqnarray}
\EuScript{I}^{2}_2 (1) &=& \frac{1}{12}(4-\eta)\,,\nonumber\\
\EuScript{I}^{3}_2 (1) &=& \frac{1}{40}(5-\eta)\,,\nonumber\\
\EuScript{I}^{4}_2 (1) &=& \frac{1}{180}(6-\eta)\,,\nonumber\\
\EuScript{I}^{1}_3 (1) &=& \frac{1}{12}(4-\eta)\,,\nonumber\\
\EuScript{I}^{2}_3 (1) &=& \frac{1}{20}(5-\eta)\,,\nonumber\\
\EuScript{I}^{3}_3 (1) &=& \frac{1}{60}(6-\eta)\,,\nonumber\\
\EuScript{I}^{4}_3 (1) &=& \frac{1}{252}(7-\eta)\,,\nonumber\\
\EuScript{I}^{2}_2 (2) &=& \frac{\pi}{24}(3-\eta)\,,\nonumber\\
\EuScript{I}^{3}_2 (2) &=& \frac{\pi}{70}(7-2\eta)\,,\nonumber\\
\EuScript{I}^{4}_2 (2) &=& \frac{\pi^2}{192}(4-\eta)\,,\nonumber\\
\EuScript{I}^{1}_3 (2) &=& \frac{1}{35}(7-2\eta)\,,\nonumber\\
\EuScript{I}^{2}_3 (2) &=& \frac{\pi}{48}(4-\eta)\,,\nonumber\\
\EuScript{I}^{3}_3 (2) &=& \frac{\pi}{126}(9-2\eta)\,,\nonumber\\
\EuScript{I}^{4}_3 (2) &=& \frac{\pi^2 }{320}(5-\eta)\,.
\label{thresholdintegralsp1}
\end{eqnarray}

\bibliography{refsReviewTM}
\end{document}